\documentclass[structabstract]{aa}
\usepackage{graphicx, amsmath, amssymb, natbib}
\usepackage{longtable}
\bibpunct{(}{)}{;}{a}{}{,}

\usepackage{txfonts}
\begin{document}
\title{Production of interstellar hydrogen peroxide (H$_2$O$_2$) on the surface of dust grains}
\titlerunning{Production of H$_2$O$_2$ on the grain}


\author{
    Fujun Du
        \inst{1}
    \thanks{Member of the International Max Planck Research School (IMPRS)
        for Astronomy and Astrophysics at the Universities of Bonn and Cologne.}
    \and
        B\'ereng\`ere Parise
        \inst{1}
    \and
        Per Bergman
        \inst{2}
}

\authorrunning{F. Du et al.}

\institute{Max-Planck-Institut f$\ddot{\rm u}$r Radioastronomie,
    Auf dem H$\ddot{\rm u}$gel 69, 53121 Bonn, Germany\\
        \email{fjdu@mpifr.de}
    \and
        Onsala Space Observatory,
               Chalmers University of Technology, 439 92 Onsala, Sweden
}

\date{}


\abstract
{The formation of water on the dust grains in the interstellar medium may
proceed with hydrogen peroxide (H$_2$O$_2$) as an intermediate. Recently
gas-phase H$_2$O$_2$ has been detected in $\rho$ Oph A with an abundance of
$\sim$10$^{-10}$ relative to H$_2$.}
{We aim to reproduce the observed abundance of H$_2$O$_2$ and other species
detected in $\rho$ Oph A quantitatively.}
{We make use of a chemical network which includes gas phase reactions as well
as processes on the grains; desorption from the grain surface through chemical
reaction is also included. We run the model for a range of physical
parameters.}
{The abundance of H$_2$O$_2$ can be best reproduced at $\sim$6$\times$10$^5$
yr, which is close to the dynamical age of $\rho$ Oph A. The abundances of
other species such as H$_2$CO, CH$_3$OH, and O$_2$ can be reasonably reproduced
also at this time. In the early time the gas-phase abundance of H$_2$O$_2$ can be
much higher than the current detected value. We predict a gas phase abundance
of O$_2$H at the same order of magnitude as H$_2$O$_2$, and an abundance of the
order of 10$^{-8}$ for gas phase water. A few other species of interest are
also discussed.}
{We demonstrate that H$_2$O$_2$ can be produced on the dust grains and released
into the gas phase through non-thermal desorption via surface exothermic reactions. The H$_2$O$_2$ molecule on the
grain is an important intermediate in the formation of water. The fact that
H$_2$O$_2$ is over-produced in the gas phase for a range of physical conditions
suggests that its destruction channel in the current gas phase network may be
incomplete.}



\maketitle


\section{Introduction}

Oxygen is the most abundant ``metal'' element in the cosmos \citep{
Savage96, Asplund09}. In the cold dense interstellar clouds, gas phase chemical models
predict that oxygen mainly resides in CO and O$_2$ molecules \citep{Herbst89,
Millar90, Wakelam06}. However, although CO is ubiquitously distributed in the interstellar
medium, O$_2$ is not. The latter is only detected very recently in $\rho$ Oph A
at a low abundance (relative to molecular hydrogen) of 5$\times$10$^{-8}$
\citep{Larsson07}, and in Orion at an abundance of (0.3 -- 7)$\times$10$^{-6}$ \citep{Goldsmith11}.  On the other hand, the observed water (gas or ice)
abundance can be as high as 10$^{-4}$ \citep{vanDishoeck04}. Thus it seems that
water, instead of O$_2$, is a main reservoir of oxygen in addition to CO.  When
only gas phase chemistry is included, the H$_2$O abundance can be of the order
10$^{-7}$ at most \citep[see, for example, ][]{Bergin00, Roberts02a} for
typical dark cloud conditions.  The fact that O$_2$ is over-produced and
H$_2$O is under-produced in gas phase chemistry suggests that adsorption onto
the grain surfaces and the reactions on the surfaces may play an
important role.

On the grain surface, H$_2$O can form through successive
additions of hydrogen atoms to an oxygen atom:
\begin{align}
   {\rm H} + {\rm O} &\rightarrow {\rm OH}, \label{eq:HO} \\
   {\rm H} + {\rm OH} &\rightarrow {\rm H}_2{\rm O}, \label{eq:HOH}
\end{align}
both of which are barrierless \citep{Allen77}. It can also form via hydrogen
addition to molecular oxygen:
\begin{align}
   {\rm H} + {\rm O}_2 &\rightarrow {\rm O}_2{\rm H} \label{eq:HO2}, \\
   {\rm H} + {\rm O}_2{\rm H} &\rightarrow {\rm H}_2{\rm O}_2 \label{eq:HHO2}, \\
   {\rm H} + {\rm H}_2{\rm O}_2 &\rightarrow {\rm H}_2{\rm O} + {\rm OH} \label{eq:HH2O2}.
\end{align}
Reaction (\ref{eq:HO2}) was assumed to have an activation barrier of 1200~K in
\citet{Tielens82}.  However, based on experimental results, \citet{Cuppen10}
recently concluded that it is barrierless.
Other possible formation pathways of water include the reaction between H$_2$
and OH and the route with O$_3$ as an intermediate \citep{Tielens82}.

In the second route described above (equations (\ref{eq:HO2} -- \ref{eq:HH2O2})),
hydrogen peroxide (HOOH, also written as H$_2$O$_2$, which is adopted in this
paper) appears as an intermediate product. Thus if this route is indeed
important, a significant amount of H$_2$O$_2$ might form on the grain, and its
gas phase counterpart could also be detectable if effective desorption
mechanisms exist.

In the current mainstream gas phase reaction networks for astrochemistry, H$_2$O$_2$ is not
efficiently formed in the gas phase.  For example, in the 2009 version of the OSU
network\footnote{http://www.physics.ohio-state.edu/$\sim$eric/research\_files/osu\_01\_2009},
the only two reactions leading to the formation of H$_2$O$_2$ are
\begin{align*}
   {\rm H}_2 + {\rm O}_2{\rm H} &\rightarrow {\rm H}_2{\rm O}_2 + {\rm H}, \\
   {\rm OH} + {\rm OH} &\rightarrow {\rm H}_2{\rm O}_2 + h\nu.
\end{align*}
The first one has a large activation barrier of 10$^4$ K, rendering it inactive
at low temperatures.  H$_2$O$_2$ is mainly consumed by
\begin{align*}
   {\rm H}_2{\rm O}_2 + h\nu &\rightarrow {\rm OH} + {\rm OH}, \\
   {\rm OH} + {\rm H}_2{\rm O}_2 &\rightarrow {\rm H}_2{\rm O} + {\rm O}_2{\rm H},
\end{align*}
the first of which is dissociation by cosmic-ray induced radiation. Other
destruction channels by reacting with H and O are ineffective due to large
activation barriers. At a temperature of 10 K and an H$_2$ density of 10$^4$
cm$^{-3}$, the steady-state abundance of H$_2$O$_2$ can be approximated by
\begin{equation*}
   X({\rm H}_2{\rm O}_2) \simeq 10^3 X^2({\rm OH}) \simeq 5\times10^{-12}.
\end{equation*}
At a higher density, the abundance of H$_2$O$_2$ will be even less because OH
is less abundant in this case.
With the UMIST RATE06 network the abundance of H$_2$O$_2$ is essentially zero
\citep{Woodall07}.

Thus if a substantial amount of H$_2$O$_2$ can be detected in the interstellar
medium, then it must have been synthesized on the dust grains, rather than in
the gas phase; this would also provide information and constraints on the
formation route of H$_2$O.

And it was indeed recently detected (for the first time) in the $\rho$ Oph A cloud by
\citet{Bergman11b}, at an abundance of $\sim$10$^{-10}$, which is well above
what would be predicted by the gas phase chemistry, indicating that chemical
processes on the grains are responsible for this detection.  Why this molecule has not been detected in the past seems to be a puzzle and will be discussed later.

In the present work we aim at modeling the gas phase abundance of H$_2$O$_2$ at
a physical condition relevant to $\rho$ Oph A, to demonstrate whether the grain
chemistry is able to explain its observed abundance. The model is also required
to give consistent abundances for other species detected earlier in this region.
Also, ice and gas-phase abundance predictions for previously undetected
species will be done.

The remaining part of this paper is organized as follows. In section
\ref{sec:chemmodel} we describe the chemical model being used in this work. In
section \ref{sec:results} we present the results of our modeling. The
conclusions are in section \ref{sec:concl}.  Appendix \ref{app:expSpike}
contains an explanation to a spike-like feature in the evolution curves of some
species.  The surface reaction network we use is listed in appendix
\ref{app:surfacenetwork}, and the enthalpies of the surface species which are
needed in the chemical desorption mechanism (see section \ref{sec:chemmodel})
are listed in appendix \ref{app:speciesenthalpy}.

\section{Chemical model}
\label{sec:chemmodel}


For the gas phase chemistry, we make use of a subset of the UMIST RATE06
network\footnote{http://udfa.net} \citep{Woodall07}.  Species containing Fe,
Na, Mg, Cl are excluded.  In total 284 gas phase species and 3075 gas phase
reactions are included. The cosmic-ray ionization rate is taken to be the
canonical value of 1.36$\times10^{-17}$ s$^{-1}$ \citep{Woodall07}.

The surface chemical network is a combination of a selection of the reactions
in \citet{Allen77}, \citet{Tielens82} and \citet{Hasegawa92}, with the rates of
a few reactions updated according to the recent experimental results and/or
theoretical calculations. In total 56 surface species and 151 surface reactions
are included (see appendix \ref{app:surfacenetwork}).

The binding energies of the surface species are either taken from
\citet{Hasegawa93a}, or estimated from the value of a similar species in a way
similar to \citet{Garrod08a}.  These values are applicable to bare grains,
i.e., grains without an ice mantle. A grain will be covered by ice (typically
water) as adsorption and reaction proceeds, thus these values are not always
appropriate. Ideally, they should be varied according to the real-time
composition of the grain. The water ice mantle mainly affects the binding
energies of species with a hydrogen bond, such as OH and H$_2$O. The effect of
this should be minor for our purpose, because most of the reactions involving a
species with a hydrogen bond are primarily mediated by another reaction partner
which does not have a hydrogen bond.

The barriers against surface diffusion are taken to be a fixed fraction of the
binding energies. A range of values have been used for this fraction in the
past, from 0.3 \citep{Hasegawa92} through 0.5 \citep{Garrod08a} to 0.77
\citep{Ruffle00}.  We use a value of 0.77, based on
the analysis of \citet{Katz99}.  Because our model is mainly for a relatively
high temperature ($\sim$20~K) in comparison with most of the previous models
(predominantly for a temperature of $\sim$10 K), a low diffusion barrier for the
surface species would lead to an unrealistic ice mantle composition. The effect
of changing this parameter is discussed later.  We allow H and H$_2$ on a dust
grain to migrate through quantum tunneling or thermal hopping, depending on
which is faster; all the heavier species are only allowed to move by thermal
hopping. The quantum tunneling and thermal hopping rates are calculated using
the formulation of \citet{Hasegawa92}. For calculation of the quantum tunneling
rates, we use a barrier width of 1 $\AA$. The exact value of this width depends
on the composition and structure of the surface, which has not been fully
quantified.

About the activation barrier of reaction (\ref{eq:HO2}), as there is a big
discrepancy between the value adopted in the past and the value proposed
recently based on experiments \citep{Cuppen10}, we adopt an intermediate value
of 600 K.  However, the effect of varying this parameter is also tested during
the modeling, and will be discussed later.  Reaction (\ref{eq:HH2O2}) has a
barrier of 1400 K in \citet{Tielens82}, while in \citet{Cuppen10} this reaction
is barrierless. We choose to adopt the latter result in this case, because too
high a barrier for it would result in building-up too much of H$_2$O$_2$ on the grain surface.

Surface reactions with an activation barrier are allowed to proceed thermally
or through quantum tunneling, depending on whichever is faster. The formula
used to calculate the rates is also the same as in \citet{Hasegawa92}, and the
reaction barriers are assumed to have a width of 1 $\AA$, although different
values are possible \citep{Garrod11}.


The reaction rate of a two-body surface reaction $${\rm A+B\rightarrow
C+\cdots}$$ is $$[k_{\rm diff}({\rm A}) + k_{\rm diff}({\rm B})]N({\rm
A})N({\rm B)}/N_{\rm S},$$ if ${\rm A\ne B}$. Here $N_{\rm S}$ is the number of
reaction sites of a grain, $k_{\rm diff}({\rm A})$ and $k_{\rm diff}(B)$ are
the diffusion rates of A and B, and $N({\rm A})$ and $N({\rm B})$ are the
number of species A and B on a single grain, respectively.  If ${\rm A=B}$,
then the reaction rate should be $$k_{\rm diff}(A)N({\rm A})(N({\rm
A})-1)/N_{\rm S}.$$  With a number density of reaction sites being 10$^{15}$
cm$^{-2}$, a dust grain with radius 0.1 $\mu$m has a $N_{\rm S}$ of about 10$^6$.

As we are mainly concerned with the gas phase abundances of several species,
their desorption mechanism must be treated carefully, especially if they are
mainly produced on the grains.  Besides the normal thermal desorption, species
can also get evaporated episodically when a cosmic-ray hits a grain.  This is
treated in the same manner as in \citet{Hasegawa93a}.  Furthermore, the
non-thermal desorption mechanism via exothermic surface reactions (for brevity we call it ``chemical desorption'' hereafter) proposed by \citet{Garrod06} \citep[see
also][]{Watson72, Garrod07, Cazaux10} is also included.  Here the products of the
exothermic reactions on the grain have a probability to be directly ejected
into the gas phase. The rate of such a desorption mechanism depends on the
exoergicity of the reaction, as well as on the desorption energy of the products. A parameter
characterizing the efficiency of this mechanism (the parameter ``$a$'' in
\citet{Garrod07}) is introduced, which we take to be 0.1. The yield of chemical
desorption is directly proportional to this ``a'' parameter, although it is not
well-constrained. The value we adopt here gives a good match to the observational results.
See section \ref{sec:limmodel} for a further discussion.
The exoergicities of these reactions are estimated from the
enthalpies of the reactants and products in the same manner as in
\citet{Allen77} (their equations (3) and (4)), and the enthalpies of the
species involved in these reactions are taken from \citet{Binnewies99} or the
NIST chemistry web book\footnote{http://webbook.nist.gov/chemistry/
\label{ft:webbook}}, or some other sporadic sources (see appendix
\ref{app:speciesenthalpy}).

However, the desorption mechanisms described above alone are not always
sufficient to provide enough gas phase abundances for some species, especially
at late times. Even if a large amount of a species is produced on the grain and
released into the gas phase at early times, later it would be accreted back to
the grain surface.  If at this later time its production is no longer active
(due to the exhaustion of the precursor species), its gas phase abundance
cannot be maintained. Dust sputtering \citep{Tielens94} and photo-desorption
\citep{Oberg07} might help to release them to the gas phase, both of which
should not be of great importance in a quiet cold dark cloud. Another possible
mechanism is that cosmic-ray induced radiation can dissociate the species on
the grain, and when the fragments recombine, the products can possibly be
ejected into the gas phase directly because of the energy release of the
reaction, as described before. We implement this mechanism in the same way as
in \citet{Ruffle01} and \citet{Garrod11} \citep[see also][]{Cuppen07}, namely,
the cosmic-ray induced photo-dissociation rates for the surface species are
taken to be the same as in the gas phase.  Several dissociation branches from
\citet{Garrod08a} (their Table 1) are included.

In our model the numbers of all the species on a single grain are solved with
the hybrid moment equation (HME) approach \citep{Du11}. It has been shown in
\citet{Du11} that the rate equation method can be inaccurate in some cases, and
the HME approach provides a major improvement over the rate equation method.
As the HME approach is relatively new, in several cases we also benchmarked our
HME results with the exact Monte Carlo method \citep[similar to the one of]
[]{Vasyunin09} as in \citet{Du11}, and the agreement is satisfactory. It is
impractical to run all the models with the Monte Carlo method because the run
time would be too long.

At present the layered structure of the grain mantle is not incorporated into
our model. Although such a structure might be more realistic, and it is
important for retaining some of the ice species, however, it is also possible
that the interstellar dust grains may have an amorphous structure, which
renders the layered structure an inaccurate description. On the other hand,
particles landing on a grain are able to penetrate into the interior by several
to tens of layers, as demonstrated by experiments \citep[see, e.g.,
][]{Ioppolo10}, thus although a model neglecting the layered structure is not
accurate, one which deactivates all the layers below the outermost surface does
not reflect the whole reality either.  In fact, for the H-addition reactions,
whether the layered structure is taken into account or not only plays a minor
role in determining the reaction rates in the accretion limit (i.e. when the
accretion and evaporation processes are much slower than the reactions
\citep{Garrod08}), because the species involved in these reactions never build up a full layer.



\section{Results and discussions}
\label{sec:results}

\subsection{Modeling $\rho$ Oph A}
\label{sec:resOphA}

\begin{figure*}[htbp]
\centering
\includegraphics[width=0.49\textwidth]{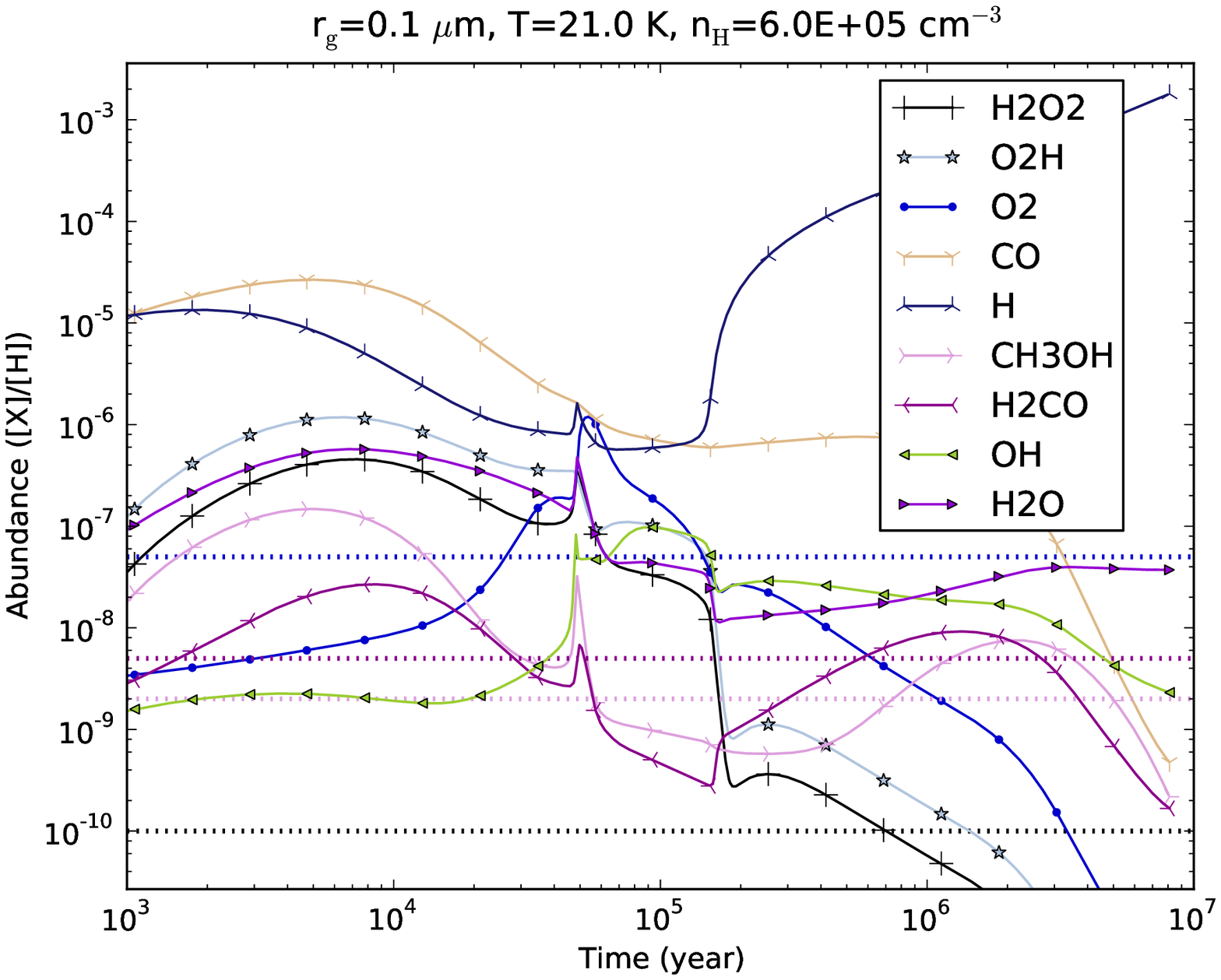}%
\includegraphics[width=0.49\textwidth]{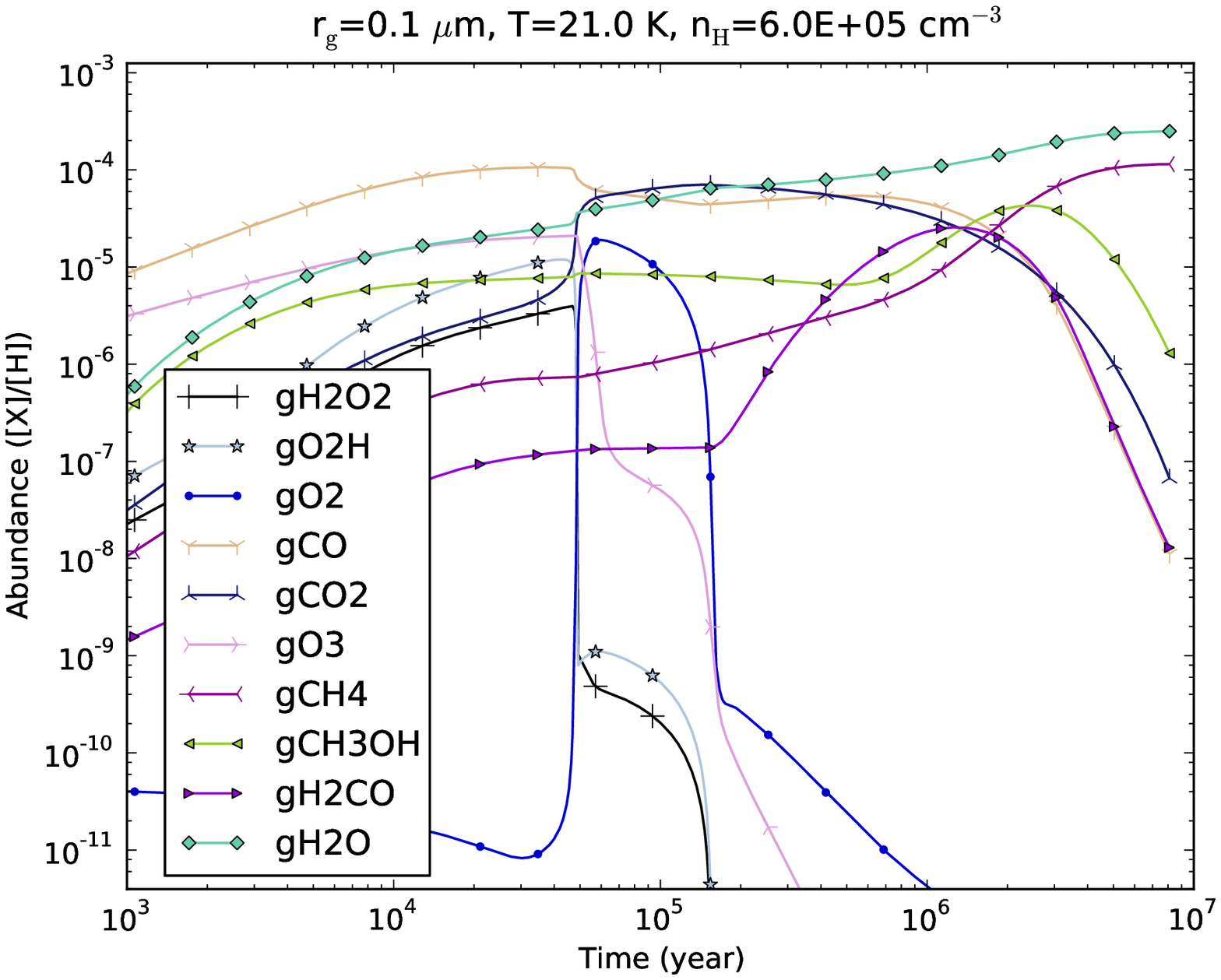}
\caption{The time evolution of the abundances of selected species. A prefix
``g'' means a grain surface species (right panel), while a name without a prefix
means a gas phase species (left panel). The dotted lines are the observed
abundances (relative to H$_2$) of O$_2$ (blue), H$_2$CO (cyan), CH$_3$OH
(magenta), and H$_2$O$_2$ (black), respectively, in the $\rho$ Oph A source.}
\label{fig:res1}
\end{figure*}

We have run the chemical model for physical parameters which are appropriate
for $\rho$ Oph A, where H$_2$O$_2$ is first detected by \citet{Bergman11b}. In
Fig.~\ref{fig:res1} we show the abundances of several species as a function of
time. In this model, a temperature of 21 K and a hydrogen density of
6$\times$10$^5$ cm$^{-3}$ has been assumed, which are determined for $\rho$ Oph
A observationally by \citet{Bergman11a,Bergman11b}. The dust
temperature and gas temperature are assumed to be the same.  A fixed value of
15 for the visual extinction $A_{\rm V}$ has been adopted. We assume a
canonical grain size of 0.1 $\mu$m and a reaction site density of 10$^{15}$
cm$^{-2}$.  The dust-to-gas mass ratio is set to 0.01, and the dust grain material is
assumed to have a mass density of 2~g~cm$^{-3}$. We also assume
that the ratio between the diffusion barrier and the binding energy is 0.77,
which is at the higher end of the values that have been used in the past, to
give a reasonable ice composition.  The initial condition is atomic except for
H$_2$. The elemental abundances are the same as in \citet{Garrod11}.

The observed abundance of gas phase H$_2$O$_2$ is $\sim$10$^{-10}$ relative to
H$_2$ \citep{Bergman11b}.  This value is best matched at a time of
$\sim$6$\times$10$^5$ yr.  In the early time, before about 2$\times$10$^5$
years, the gas phase H$_2$O$_2$ abundance can be as high as
$\sim$5$\times$10$^{-7}$. At late times the H$_2$O$_2$ abundance decreases to a
very low value, due to the exhaustion of O$_2$ on the grain and a full conversion of H$_2$O$_2$ into H$_2$O.

H$_2$O$_2$ is mainly formed through reaction (\ref{eq:HHO2}) on the grain,
followed by immediate desorption of the product into the gas phase caused by
the reaction heat. About 7\% of the produced H$_2$O$_2$ are released this way.
Its gas phase abundance is determined by the adsorption and chemical desorption
processes. The dissociation of gas phase H$_2$O$_2$ by cosmic-ray-induced
photons is unimportant in consuming it, in comparison with adsorption.

\citet{Ioppolo08} modeled the abundance of H$_2$O$_2$ ice briefly, giving a
value of 10$^{-14}$ -- 10$^{-10}$ relative to molecular hydrogen, depending on
which energy barriers of several relevant reactions have been used.  Our major
goal is to model the gas phase H$_2$O$_2$ abundance, rather than the H$_2$O$_2$
ice.  In our model results, the gas phase abundance of H$_2$O$_2$ is much
higher than its ice counterpart.  The H$_2$O$_2$ ice does not have a
significant abundance in the later stage, being well below the upper limit
(5.2\% with respect to H$_2$O ice) given by \citet{Boudin98}, because it is
constantly transformed into H$_2$O by reacting with the accreted H atoms.  However, by
irradiating thin water ice film with low energy ions, \citet{Gomis04} found
that it is possible to obtain an H$_2$O$_2$ to H$_2$O ratio in the solid phase
up to a few percent. This direct processing of the grain mantle by cosmic rays
is not included in our model.  The production of H$_2$O$_2$ inside water ice in
an O$_2$ rich environment triggered by UV radiation \citep{Shi11} should also
be of little importance here.  We notice in Fig.~\ref{fig:res1} (right panel)
that in the early stage (before $\sim$5$\times$10$^4$ yr) the H$_2$O$_2$ ice
can achieve a rather high abundance, $\sim$5$\times$10$^{-6}$ relative to H
nucleus or $\sim$10\% relative to H$_2$O.  During this early period the water
formation on the grain mainly proceeds through reaction (\ref{eq:HH2O2}) with
H$_2$O$_2$ as an intermediate, which is responsible for about half of the final
water ice repository on the grain mantle. In the later stage reaction
(\ref{eq:HOH}) takes over.  In the results of \citet{Ioppolo08} (their Fig.~4)
we do not see a similar feature (i.e. a high abundance of H$_2$O$_2$ ice in the
early stage). In our current model the layered structure of the grain mantle is
not taken into account. It is possible that, if such a structure is considered,
the inner layers with a relatively high H$_2$O$_2$ content might be maintained,
which would give a value of a few percent for the H$_2$O$_2$ to H$_2$O ratio in
the solid phase.


Methanol (CH$_3$OH) and formaldehyde (H$_2$CO) are also detected in the $\rho$
Oph A SM1 core \citep{Bergman11a}, at an abundance of $\sim$2$\times$10$^{-9}$
and $\sim$5$\times$10$^{-9}$, respectively.  Their abundances are also
reproduced very well at a time of $\sim$6$\times$10$^5$ yr in our model.  At
early times, both CH$_3$OH and H$_2$CO have a high abundance.  Their
abundances also have a peak in the period between 2$\times$10$^5$ yr to 10$^7$ yr.  In
our current network, CH$_3$OH is mainly formed on the grains, and mainly
through the addition of H atom to CH$_2$OH, while the latter is mainly produced from
the reaction between C and OH to form HOC followed by successive H additions.
Thus the abundance of CH$_3$OH decreases at very late times due to the
depletion of atomic C (which is mainly in CH$_4$ ice in the late stage).  The
normal formation channel through successive hydrogenation of CO is important
at around 5$\times$10$^5$ -- 2$\times$10$^6$ yr.  The gas phase H$_2$CO mainly forms in the
gas phase in the early stage ($<10^5$ yr), and mainly through the reaction
${\rm CH}_3+{\rm O}\rightarrow{\rm H}_2{\rm CO}+{\rm H}$.  Later it is mainly
formed through successive hydrogenation of CO on the grain surface followed by
chemical desorption. The abundance of methanol and formaldehyde ice relative
to water ice can be as high as $\sim$20\% at their peaks at a time of
$\sim$2$\times$10$^6$ yr, but falls down to a very small value in the late
times.  The late time abundances are consistent with the upper limit derived
for quiescent environment and low mass young stellar objects in
\citet{Gibb04}.  However, in \citet{Pontoppidan04} a much higher abundance of
CH$_3$OH ice is observed along the line-of-sight of SVS4 (a dense cluster of
pre-main sequence stars) which is close to a class 0 protostar; this is
consistent with the peak abundances in our model.

From Fig.~\ref{fig:res1} (left panel) it can be seen that the abundance of
gaseous O$_2$ at an intermediate time of 6$\times$10$^5$ year is
$\sim$6$\times$10$^{-9}$, which is within one order of magnitude of the
observed abundances of 5$\times$10$^{-8}$ for O$_2$ \citep{Larsson07}. The
late-time abundance of O$_2$ drops to a very low value, while its observed
abundance is best matched at a time of $\sim$2$\times$10$^5$ yr. We notice that
during the period $\sim$(0.6 -- 2)$\times$10$^5$ yr, the abundance of O$_2$ ice
has a prominent bump, reaching a peak abundance of $\sim$10$^{-5}$ relative to
H$_2$. At the same time, the gas phase O$_2$ also reaches an abundance of
$\sim$(1 -- 5)$\times$10$^{-7}$. These values can be compared with the recent
detection of gas phase O$_2$ at an abundance of (0.3 -- 7)$\times$10$^{-6}$ in
Orion by Herschel \citep{Goldsmith11}. Warm-up of the dust grain at this stage
may release a large amount of O$_2$ molecule into the gas phase.

As a precursor of H$_2$O$_2$, O$_2$H mainly forms from the reaction between O
and OH on the grain, which does not have a barrier according to
\citet{Hasegawa92}.  The ratio between the gas phase O$_2$H and H$_2$O$_2$ is
almost constant throughout the evolution, being approximately 3.  Thus in our
current network the gas phase  O$_2$H also has a remarkable abundance, which
might be detectable in the future.

Except at very early times, the grain mantle is mainly composed of water ice.  The abundances of CO and
CO$_2$ are comparable at an intermediate time of (0.3--1)$\times$10$^6$ yr,
being about 40--60\% of water ice.  This is in rough agreement with the ice
composition for intermediate-mass YSOs \citep{Gibb04} \citep[see
also][]{Oberg11}, and is also in line with the suggestion of \citet{An11} that
CO$_2$ ice is mixed with CH$_3$OH ice (the latter is about 10\% of the former
in our model).  Water ice mainly forms from reaction (\ref{eq:HH2O2}) in the
early time, and from reaction (\ref{eq:HOH}) in the late time; the gas phase
formation route of water only plays a minor role. On the contrary, the CO ice
mantle is mainly from accretion of CO molecules formed in the gas phase.  For
the CO$_2$ ice, it is mainly accreted from its gas phase counterpart in the
early time, and its abundance is further increased through the reaction ${\rm
OH}+{\rm CO}\rightarrow{\rm CO}_2+{\rm H}$ in the late stage. At late times
($>3$$\times$10$^6$ yr), most of carbon resides in the form of CH$_4$ ice, the
latter being about half of the water ice. However, as far as we know, such a
high abundance of CH$_4$ ice \citep[see also][]{Garrod11} has not been observed
in the interstellar medium.

The gas phase CO is heavily depleted, with an abundance $\sim$10$^{-6}$
relative to H nucleus or $\sim$1\% relative to its ice counterpart, at an
intermediate time (10$^5$ -- 10$^6$ yr). Its abundance is mainly determined by
the balance between the adsorption and cosmic-ray induced evaporation
processes. At late times the CO ice abundance drops to a very low value. This
is because CO is continuously hydrogenated into H$_2$CO or CH$_3$OH, and the
dissociation of CH$_3$OH by cosmic-ray induced photons produces CH$_3$, which
quickly becomes CH$_4$ by hydrogenation. Taking into account the layered
structure of grain mantle can retain a fair amount of CO ice.

The abundance of gas phase H$_2$O in the intermediate to late time is of the
order 10$^{-8}$. At these times, its grain surface production route
(\ref{eq:HOH}) followed by chemical desorption and the gas phase production
route ${\rm H}_3{\rm O}^+ +{\rm e}^-\rightarrow{\rm H}_2{\rm O}+{\rm H}$ plays
a similarly important role. H$_3$O$^+$ itself is mainly formed from successive
protonation of atomic oxygen at this stage.

The hydroxyl radical (OH) has an abundance $\sim$10$^{-8}$ -- 10$^{-9}$ in the
gas phase at the intermediate to late times.  These values are comparable to
the observed abundance of $\sim$(0.5 -- 1)$\times$10$^{-8}$ in the envelope
around the high-mass star-forming region W3 IRS 5 obtained with HIFI on board
{\it Herschel} \citep{Wampfler11}.  Although the physical condition in this
source is different from $\rho$ Oph A, however, as OH is readily produced and
recycled in the gas phase by the reactions ${\rm H_3O^++e^-\rightarrow OH+2H}$
and ${\rm H_3^++OH\rightarrow H_2O^++H_2}$, grain processes will not play a
dominant role in determining its abundance, especially when the temperature is
not too low.  However, \citet{Goicoechea06} detected a much higher abundance
(0.5 -- 1)$\times$10$^{-6}$ of OH in the Orion KL outflows, which seems to
require other formation pathways (e.g. shock destruction of H$_2$O ice).

A relatively high abundance ($\sim$10$^{-7}$ and $\sim$10$^{-5}$) of gas phase
and grain surface ozone (O$_3$) is also obtained for $t\lesssim10^5$ yr.  However,
these values should be treated with caution, because the chemistry of O$_3$ is
very incomplete in our current model: neither the OSU gas
phase network nor the UMIST RATE06 network includes it. Possible gas phase destruction
pathways involving atomic O and S (according to the NIST chemistry web book;
see footnote \ref{ft:webbook}) may lower its gas phase abundance significantly,
but its ice mantle abundance should not be severely affected.



The abundance of atomic hydrogen in the gas phase is quite high at
late times as seen from Fig.~\ref{fig:res1}, which seems to be at odds with the
usual results. In many gas phase chemical models, the abundance of atomic
hydrogen in the gas phase is determined by the balance between its adsorption
onto the dust grains and the dissociation of H$_2$ molecules by cosmic rays.
The adsorption process is thought to have a sticking coefficient close to
unity, and evaporation is assumed not to occur (which is appropriate at low
temperatures). In such a framework it can be found that the gas phase atomic
hydrogen will always have a fixed density of the order 1 cm$^{-3}$. However,
in our case with a temperature of 21 K, evaporation is very fast and cannot be
neglected.  Furthermore, the dissociation of ice mantle by cosmic-ray induced
photons generates atomic hydrogen, which enhances its gas phase abundance
significantly in the late stage.

We note that at a time of about 5$\times$10$^4$ years the abundances of
several species change very quickly, and the abundances of some other species
show a spike-like feature. This may resemble at first sight an erroneous
behavior caused by the differential equation solver. So we ran the model with
the same parameters using a Monte Carlo code (also used for benchmark purpose
in \citet{Du11}) which is free from such problems, and it turns out that these
features are genuine.  A semi-quantitative explanation of this feature is in
appendix \ref{app:expSpike}.




\subsection{Chemical age versus dynamical time scale}
\label{sec:chemage}

As noted before, the time of best agreement between our modeling results and
the observational results of \citet{Bergman11a} and \citet{Bergman11b} is at
$\sim$6$\times$10$^5$ year. Interestingly, this time scale is quite close to
the time scales derived in \citet{Andre07} (their Table 7). For example, the
evolution time scale for $\rho$ Oph A as estimated to be three times the
free-fall time is (0.5 -- 2)$\times$10$^5$ years, being close to the
statistically estimated age, while the collisional time scale of
5.5$\times$10$^5$ years and the cross time scale of 8$\times$10$^5$ years are
also of the same order of magnitude.

It is important to define the starting point when talking about age. In the
chemical evolution model described above, the whole system starts to evolve
from a simple initial state: all the elements except for hydrogen are in atomic
form, and the grains are bare. How relevant is such an initial condition when
we talk about the dynamical evolution of a cloud condensation?  At a density of
$\sim$10$^3$ cm$^{-3}$, a temperature of $\sim$20 K, and a visual extinction of
2, which are typical of diffuse or translucent molecular clouds, a chemical
model starting from an atomic initial condition (except for H$_2$) reaches
steady-state for most of the species in several 10$^3$ years.  Because such a
time scale is much shorter than the dynamical time scale of a cloud, it should
not make much difference for a time-dependent model (i.e. one in which
temperature and density etc. vary with time) to use either atomic or molecular
initial conditions. For a time-independent cloud model (such as the one we are
using) with constant physical conditions, adopting an atomic initial condition
and a high density is equivalent to assuming that the cloud was compressed from
the diffuse interstellar medium very quickly.  Converging flows in the
interstellar medium might play such a role, although this seems unlikely in
$\rho$ Oph A because only very small velocity gradient was observed
\citep{Andre07}.

However, the time of best match and the predicted abundances may also be very
dependent on the value adopted for some of the modeling parameters that are not
very well constrained.  It is necessary to see how the results would be changed
if these parameters are varied.

\subsection{Effects of changing the energy barrier of the surface reaction
    ${\rm H} + {\rm O}_2 \rightarrow {\rm HO}_2$}
\label{sec:reacen}

The activation barrier of reaction (\ref{eq:HO2}) was set to 1200 K in
\citet{Tielens82}, which was taken from the theoretical calculation by
\citet{Melius79} for the gas phase case.  \citet{Cuppen10} concluded that this
reaction has a negligible barrier. We choose to use an
intermediate value for the energy barrier of reaction (\ref{eq:HO2}), namely,
600~K for the modeling.  Here we study how would the uncertainties in this
parameter affect the abundances of several species of interest.

\begin{figure}[htbp]
\centering
\includegraphics[width=0.49\textwidth]{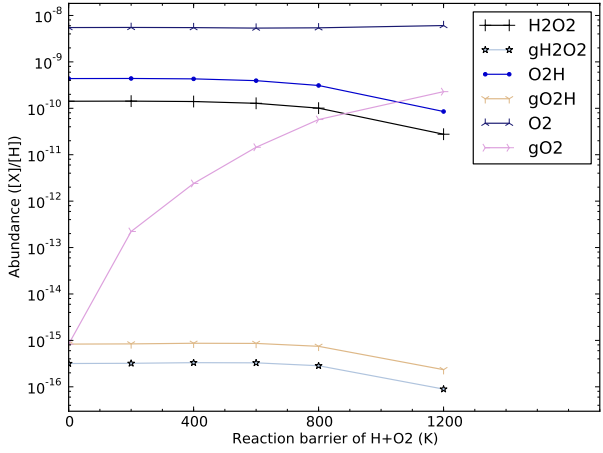}
\caption{Dependence of the abundances of several species at the best-match time
(6$\times$10$^5$ yr) on the reaction barrier of reaction (\ref{eq:HO2}), with
density fixed to 6$\times$10$^5$ cm$^{-3}$ and temperature fixed to 21 K.}
\label{fig:vsHO2}
\end{figure}

In Fig.~\ref{fig:vsHO2} the abundances of several species at the best-match
time (6$\times$10$^5$ yr) are plotted as a function of the activation energy
barrier of reaction (\ref{eq:HO2}).  The temperature and density are fixed at
21 K and 6$\times$10$^5$ cm$^{-3}$, and the ratio between the diffusion energy
barrier and the binding energy is set to 0.77.

The abundance of gas phase O$_2$ is not affected by changing the energy barrier
of reaction (\ref{eq:HO2}) because its abundance is mainly determined by the
gas phase production process ${\rm O + OH\rightarrow O_2+H}$ and adsorption.
However, the abundance of grain surface O$_2$ increases by five orders of
magnitude as the barrier changes from 0 to 1200 K, because reaction
(\ref{eq:HO2}) is one of the main reactions for the consumption of grain O$_2$.
However, even with a barrier of 1200 K for reaction (\ref{eq:HO2}), the abundance of O$_2$ on the grain at late stage is too low to be detected.
Although the rate coefficient of reaction (\ref{eq:HO2}) is reduced by 6 orders
of magnitude when its barrier changes from 0 to 1200K, the abundance of O$_2$H
on the grain does not change significantly. The reason is that when reaction
(\ref{eq:HO2}) becomes slower, more O$_2$ will build up on the grain,
compensating for the effect of increasing the barrier of reaction
(\ref{eq:HO2}).  The abundance of gas phase H$_2$O$_2$ does not change much
either, as its abundance is mainly determined by accretion onto the grain, and
production by hydrogenation of O$_2$H followed by partial chemical desorption.

\subsection{Effects of changing the diffusion energy barriers}
\label{sec:diffen}

The diffusion energy barrier of a species on the grain determines how fast it
can migrate on the grain, so it basically determines the pace of the grain
chemistry. Usually it is set to a fixed fraction (here we denote it by $\eta$)
of the binding energy of each species. The latter determines how
fast a species evaporates into the gas phase. However, this parameter might
depend on the material and morphology of the dust grain, as well as the
property of the adsorbate itself, so it is very uncertain. Values in the range
of 0.3 -- 0.77 have been adopted in the literature. We mainly used 0.77 for our
modeling. Here we investigate how different values of $\eta$
would affect the abundances of several species of interest.

\begin{figure}[htbp]
\centering
\includegraphics[width=0.49\textwidth]{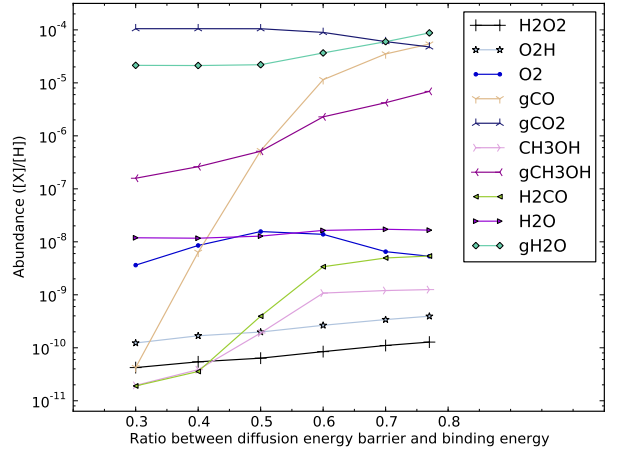}
\caption{Dependence of the abundances of several species at the best-match time
(6$\times$10$^5$ yr) on the ratio between the diffusion and desorption energy,
with density fixed to 6$\times$10$^5$~cm$^{-3}$ and temperature fixed
to 21~K.}
\label{fig:vsdiffdesor}
\end{figure}

In Fig.~\ref{fig:vsdiffdesor} we plot the abundances of several species at the
time of best-match (6$\times$10$^5$ yr) as a function of the ratio between the diffusion barrier and binding energy ($\eta$).
Changing this parameter has a large effect on the abundances of some species.
For example, the abundance of CO$_2$ ice is reduced with a higher $\eta$. This
is because at the time of concern it mainly forms through the reaction ${\rm OH
+ CO\rightarrow CO_2 + H}$, which requires the migration of two relatively
heavy species. With a low diffusion energy and a moderate temperature ($\sim$20
K), OH and CO can thermally hop quite fast, leading to a high abundance of
CO$_2$, which is not observed. However, if a lower temperature ($\sim$10 K) is
adopted, the problem becomes the opposite: the mobilities of OH and CO
are so low that it is difficult for them to meet each other to form enough
CO$_2$ ice, and certain intricate mechanism (e.g. three body reaction) has to be introduced to account for
this \citep{Garrod11}.

The abundance of H$_2$O ice increases as $\eta$ increases. At this stage it is
mainly formed by hydrogenation of OH. As the mobility of atomic hydrogen on the
grain is not greatly affected by the value of the diffusion energy barrier
because it is allowed to migrate through quantum tunneling, the reaction rate
of ${\rm H+OH\rightarrow H_2O}$ is not significantly affected by $\eta$;
however, a larger $\eta$ leaves more OH available for water because a lower
amount of it is consumed in forming CO$_2$. This is also the reason for a
higher CO (and species dependent on it such as H$_2$CO and CH$_3$OH) ice abundance
when $\eta$ is larger.  The abundances of gas phase H$_2$O$_2$ and O$_2$H also
increase for a larger $\eta$, albeit only mildly.

\subsection{Dependence on the temperature and density}
\label{sec:TandDen}

\begin{figure*}[htbp]
\centering
\includegraphics[width=0.49\textwidth]{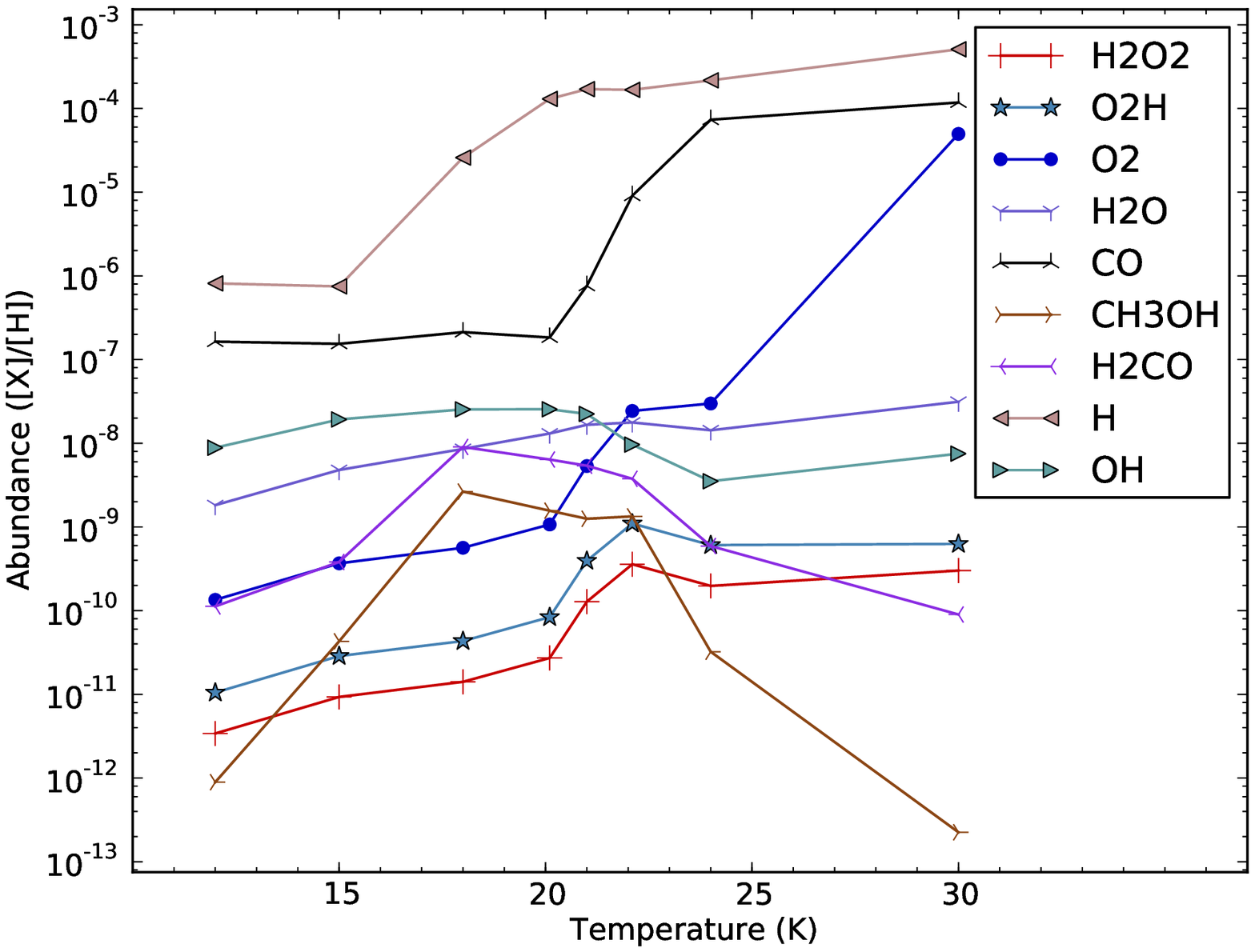}%
\includegraphics[width=0.49\textwidth]{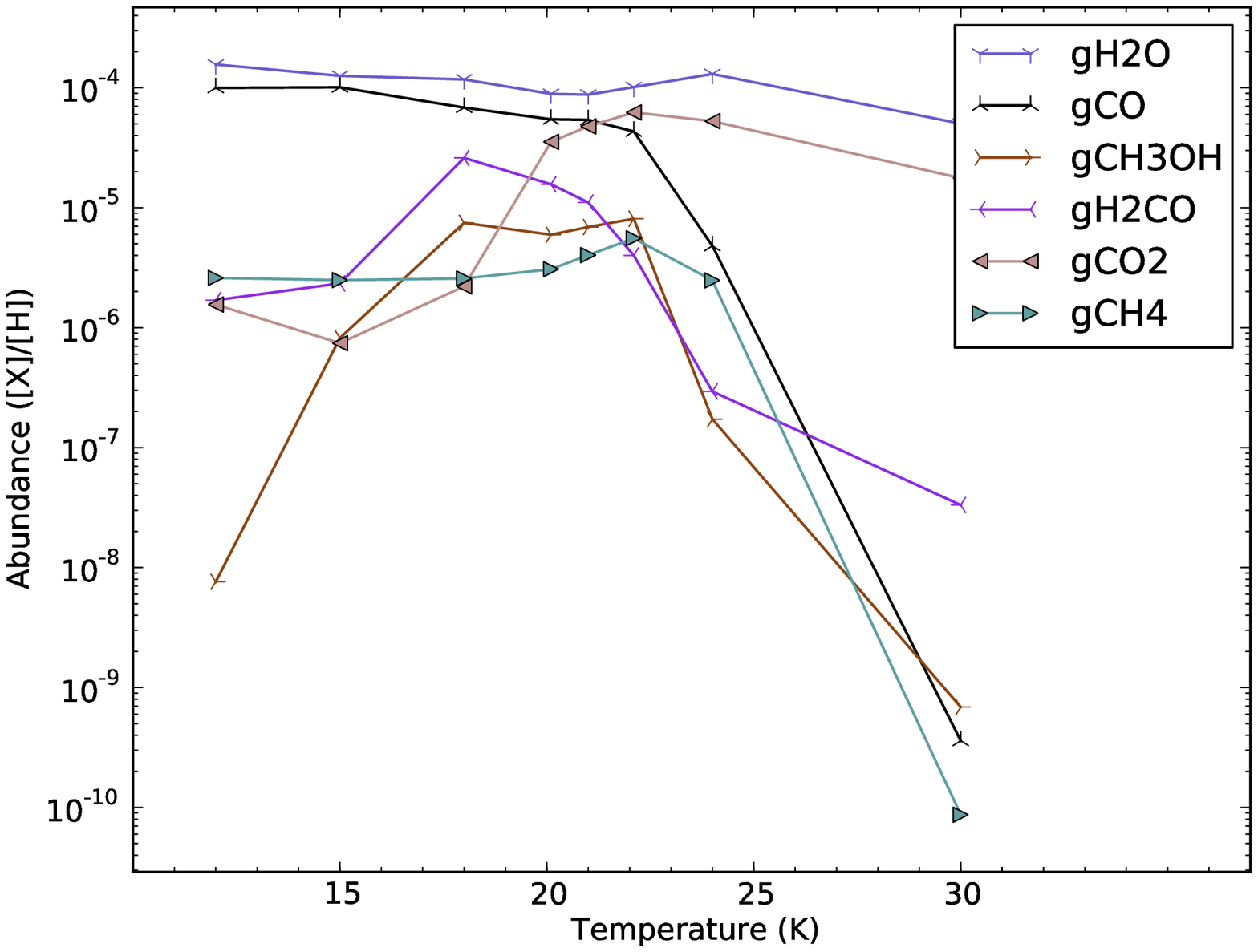}
\caption{Dependence of the abundances of several species at the time of best-match
(6$\times$10$^5$ yr) on temperature, with density fixed to 6$\times$10$^5$
cm$^{-3}$. Left panel: gas phase species; right panel: surface species.}
\label{fig:vsT}
\end{figure*}

The physical conditions (temperature and density) of $\rho$ Oph A are determined
by non-LTE radiative transfer modeling
\citep{Bergman11a}, which is usually subject to uncertainties from
many aspects, such as the excitation condition, source geometry, beam filling
factor, etc.  In this section we study how the uncertainties in the temperature
and density of the system would affect the abundances of several species in our
model.  In Fig.~\ref{fig:vsT} and Fig.~\ref{fig:vsnH} we plot the abundances of
several species at the time of best-match (6$\times$10$^5$ yr) as a function of
temperature and density.

Apparently, temperature plays a much more drastic role than density.  This is
intuitively easy to understand because temperature enters the calculation of
rates exponentially for the surface reactions.  The general trend is that when
the temperature is either too low or too high, the grain surface chemistry
tends to be inactive or unimportant. In the former case the mobilities of
species other than atomic hydrogen (which migrates through quantum tunneling in
our present model) are low, while in the latter case the surface abundances of
many species are low due to elevated evaporation rates.

As can be seen in Fig.~\ref{fig:vsT}, the abundance of CO ice starts to decrease at a temperature of around 20 K.
This value can be estimated as the temperature at which the gas phase and grain
surface abundance of CO are equal \citep[see also][]{Tielens82}, when only the
adsorption and evaporation processes are taken into account \citep[see
also][]{Hollenbach09}:
\begin{align}
  T_{\rm evap}
  &= E_{\rm D}/\ln\left[\frac{\nu}{n_{\rm H}\;{R_{\rm G}}}\frac{1}{\pi
  r^2}\frac{1}{\sqrt{8kT_{\rm gas}/\pi m}}\right]
    \\
  &\simeq E_{\rm D}/\left\{60+ \ln\left[ \left(\frac{10^5\ {\rm cm}^{-3}}{n_{\rm H}}\right)
   \left(\frac{20\ {\rm K}}{T_{\rm gas}}\right)^{1/2} \left(\frac{m}{28\
   {\rm au}}\right)^{1/2} \right]\right\}, \nonumber
\end{align}
where $E_{\rm D}$ is the
evaporation energy barrier of a species on the grain surface,
$\nu$ is the vibrational frequency of a species on the grain, ${R_{\rm G}}$ is
the dust-to-gas number ratio, $r$ is the grain radius, while $m$
is the molecular mass of the species being considered.  A
typical value of 10$^{12}$ s$^{-1}$ for $\nu$, a dust grain radius of 0.1
$\mu$m, and a dust-to-gas ratio of 2.8$\times$10$^{-12}$ have been adopted in
deriving the number 60.  Since the logarithmic part in this equation is usually
small for a typical gas density and temperature, the evaporation temperature can
be approximated simply by $T_{\rm evap}\simeq E_{\rm D}/60$.  For CO, a
canonical value of $E_{\rm D}$ is 1210 K \citep{Allen77}, which gives an evaporation
temperature of 20 K, while for water, an $E_{\rm D}$ of 1860 K
\citep{Hasegawa93a} on bare graphite grains gives an evaporation temperature of $\sim$30 K.  In
\citet{Garrod06b} a much higher desorption energy of 5700 K for water is used
(appropriate for water ice), which gives a evaporation temperature of $\sim$95
K, close to the observed evaporation temperature of water in envelope
surrounding protostars \citep{Maret02}. The evaporation temperature of CH$_4$
ice is close to that of CO ice. The evaporation time scale at the evaporation
temperature can be estimated to be roughly $\nu^{-1}\exp(E_{\rm D}/T_{\rm
evap})\simeq10^{-12}e^{60}\ {\rm s}\simeq3.6\times10^6$ yr. For higher
temperature the evaporation will be much faster.

The abundance of CO$_2$ ice initially increases with increasing temperature.
This is because it mainly forms from the reaction ${\rm CO}+{\rm
OH}\rightarrow{\rm CO}_2+{\rm H}$ with a barrier of 80 K, and an increase in
temperature greatly enhances the mobility of the reacting species, as well as
the probability of overcoming the reaction barrier.  But when temperature
increases more, the abundance of CO ice becomes so low that the CO$_2$
abundance also drops (the evaporation temperature of CO$_2$ is about 40~K
thus evaporation is not responsible for the decline in CO$_2$ ice abundance seen
in Fig.  \ref{fig:vsT}). A similar trend is seen in other species, such as
H$_2$CO ice, CH$_3$OH ice, as well as gas phase H$_2$O$_2$, CH$_3$OH, H$_2$CO,
and H$_2$O.  On the other hand, for species efficiently produced in the gas
phase, e.g.  O$_2$, its abundance increases with temperature due to the faster
evaporation at a higher temperature.

\begin{figure*}[htbp]
\centering
\includegraphics[width=0.49\textwidth]{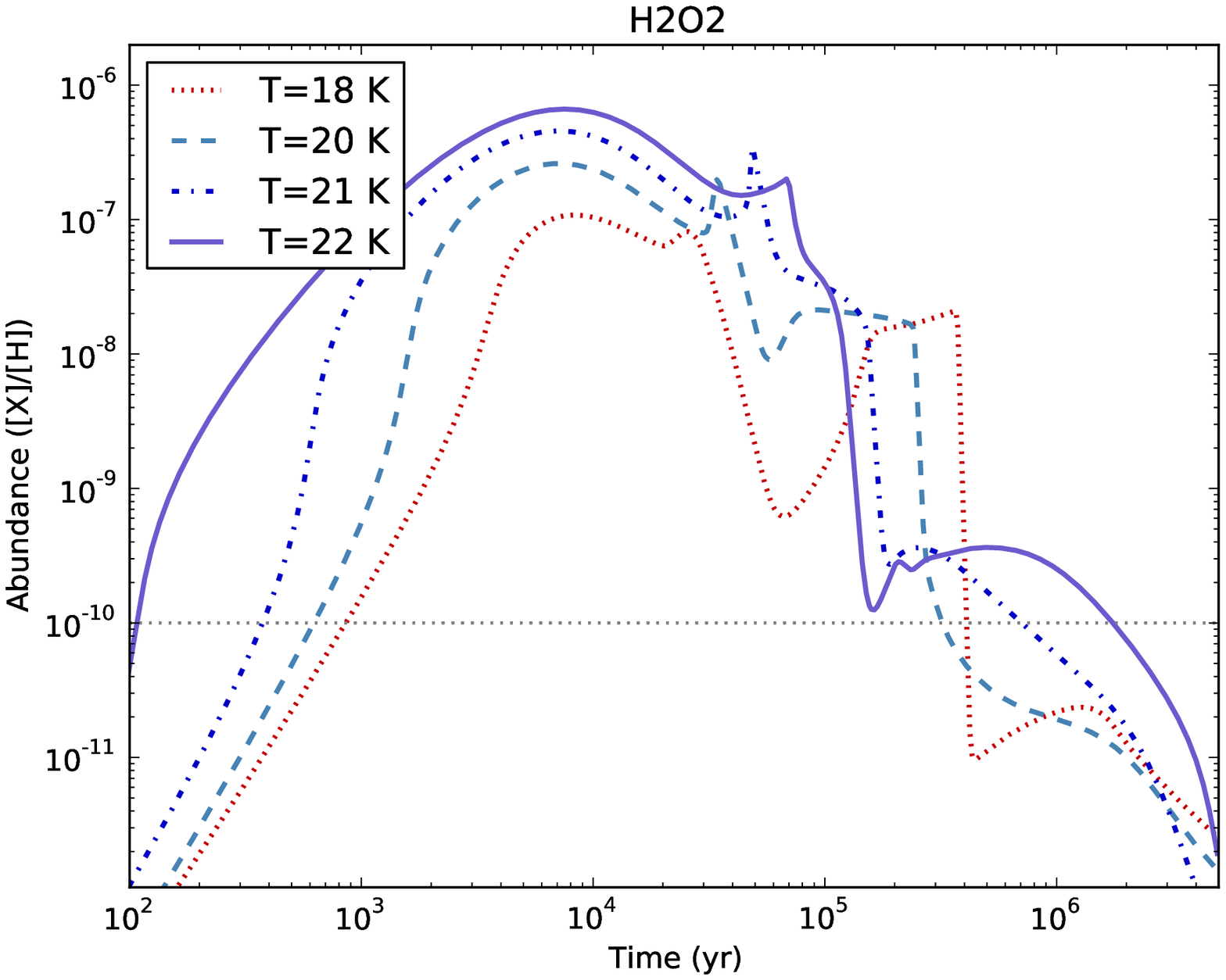}%
\includegraphics[width=0.49\textwidth]{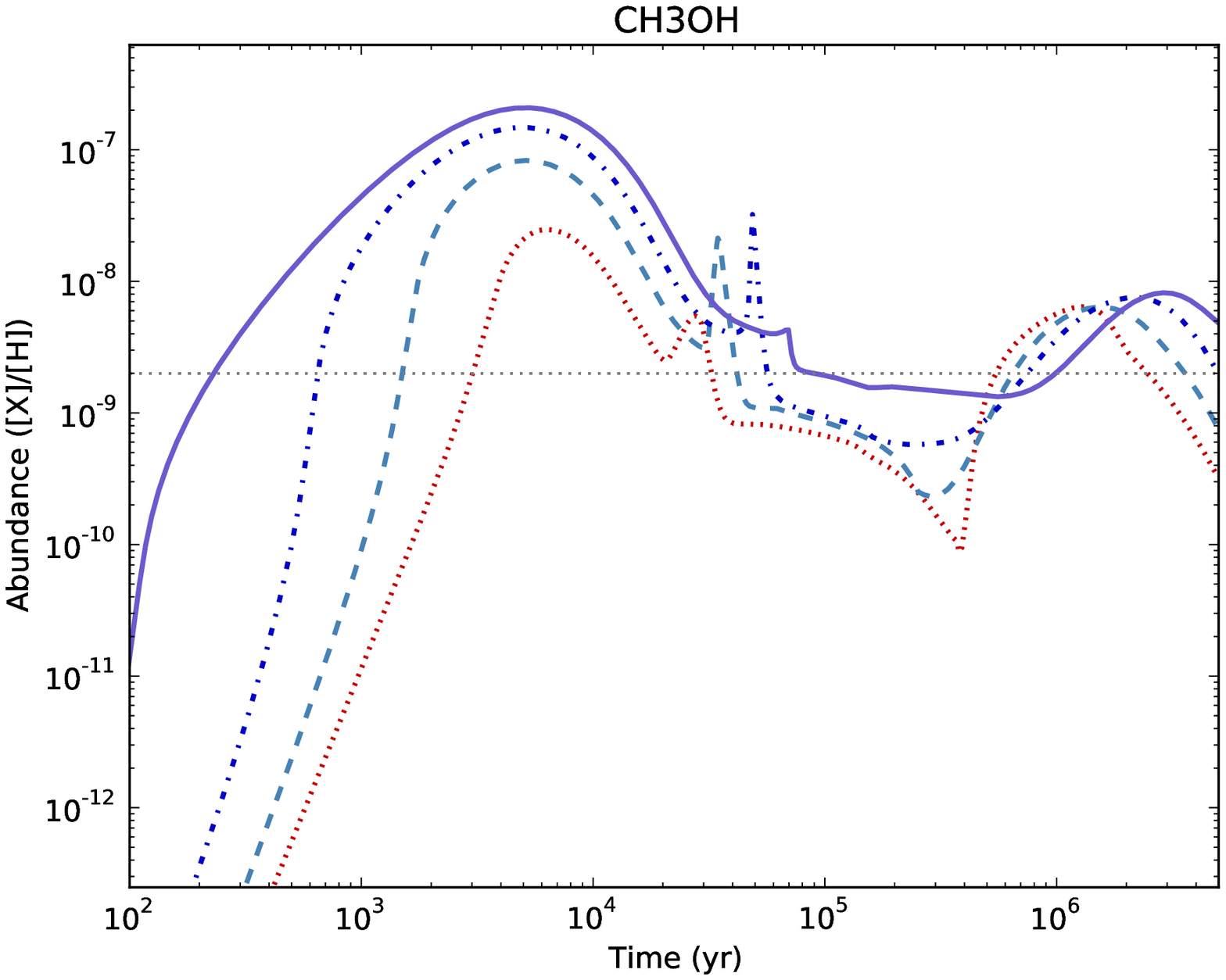}
\caption{The abundances of gas phase H$_2$O$_2$ and CH$_3$OH as a function of
time for different temperatures. The density is fixed to 6$\times$10$^5$
cm$^{-3}$. The horizontal dotted lines mark the observed abundances of the
corresponding species.}
\label{fig:togetherT}
\end{figure*}

\begin{figure*}[htbp]
\centering
\includegraphics[width=0.49\textwidth]{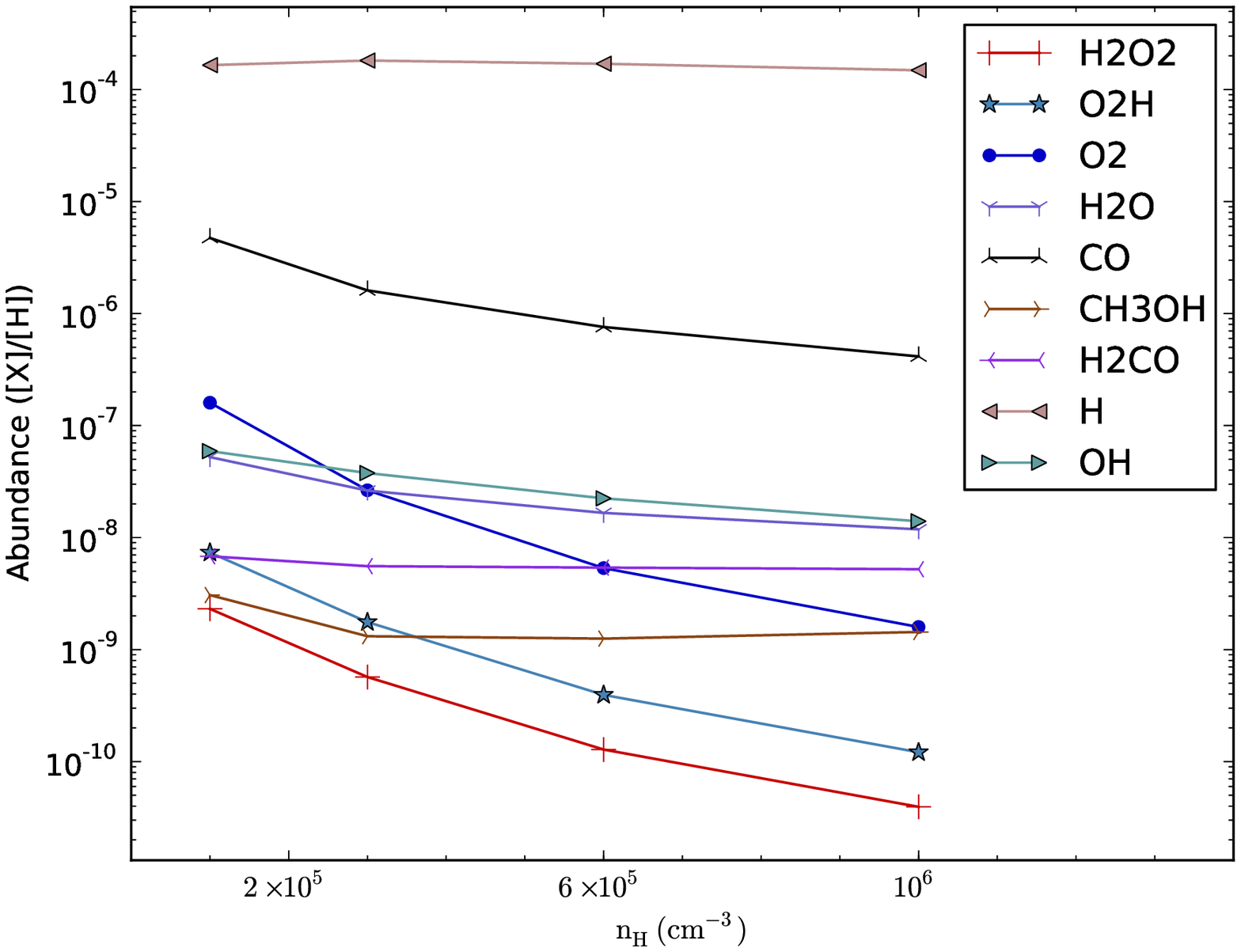}%
\includegraphics[width=0.49\textwidth]{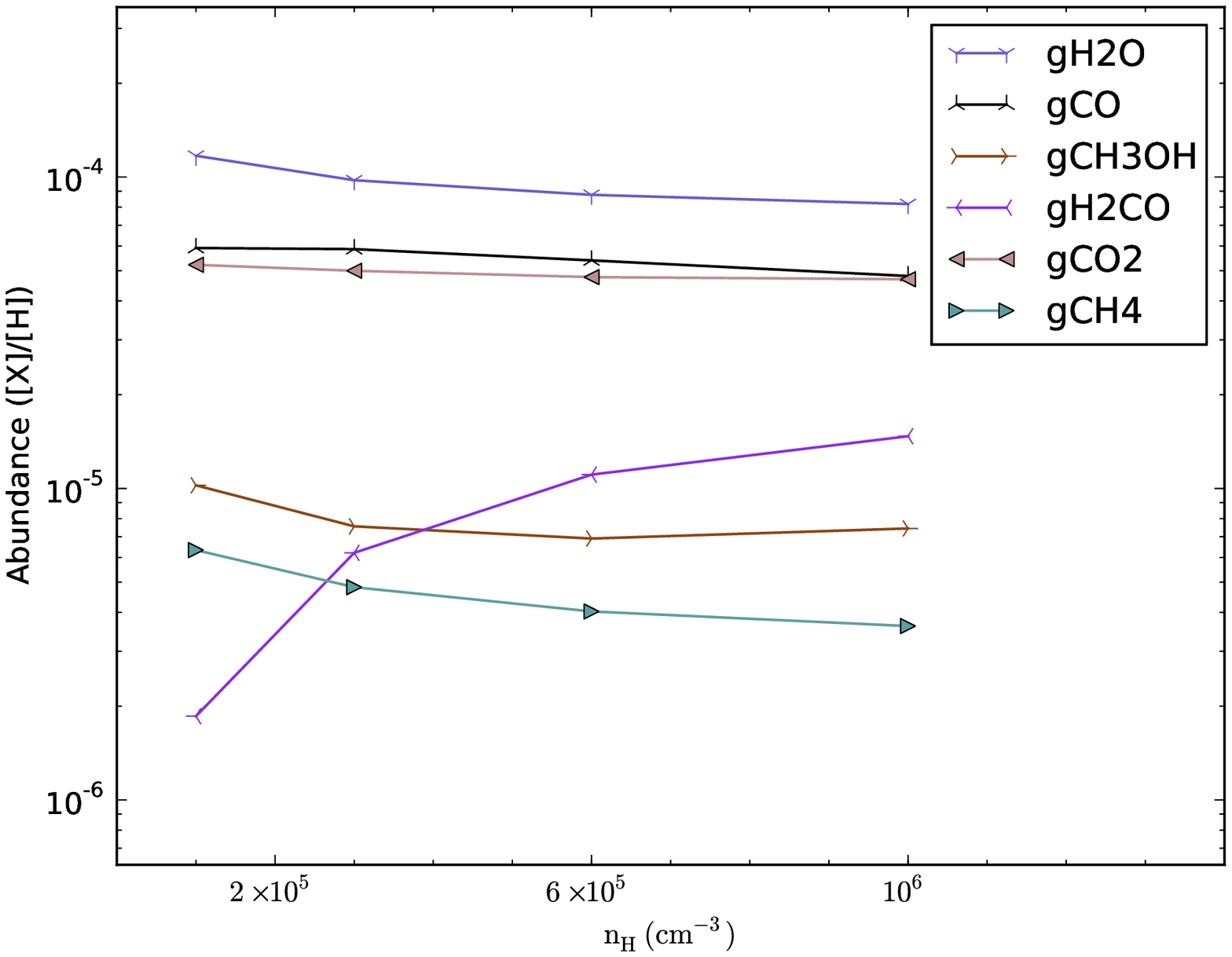}
\caption{Dependence of the abundances of several species at the time of
best-match (6$\times$10$^5$ yr) on n$_{\rm H}$, with temperature fixed to 21 K.
Left panel: gas phase species; right panel: surface species.}
\label{fig:vsnH}
\end{figure*}

The abundances of gas phase H$_2$O$_2$, CH$_3$OH, CO, and O$_2$ have quite a
sensitive dependence on temperature at around 21 -- 24~K as seen from Fig.~\ref{fig:vsT}.  For
example, changing the temperature from 20 K to 22 K increases the abundance of
H$_2$O$_2$ at a given time (6$\times$10$^5$ yr) by about one order of
magnitude.
This rather small change of 2 K in temperature is normally smaller than the
accuracy of the kinetic temperature as determined from observational data.
The dependence of the evolution curves of H$_2$O$_2$
and CH$_3$OH on temperature can be seen more clearly in
Fig.~\ref{fig:togetherT}.  Changing the temperature not only shifts the
evolution curves horizontally, it also changes their shapes significantly.  We
can see that although for CH$_3$OH it is possible to match the observed
abundance at multiple stages, for H$_2$O$_2$ the best match is only possible
at 3$\times$10$^5$ -- 10$^6$ yr (if we ignore the match in the very early
stage).

\begin{figure*}[htbp]
\centering
\includegraphics[width=0.49\textwidth]{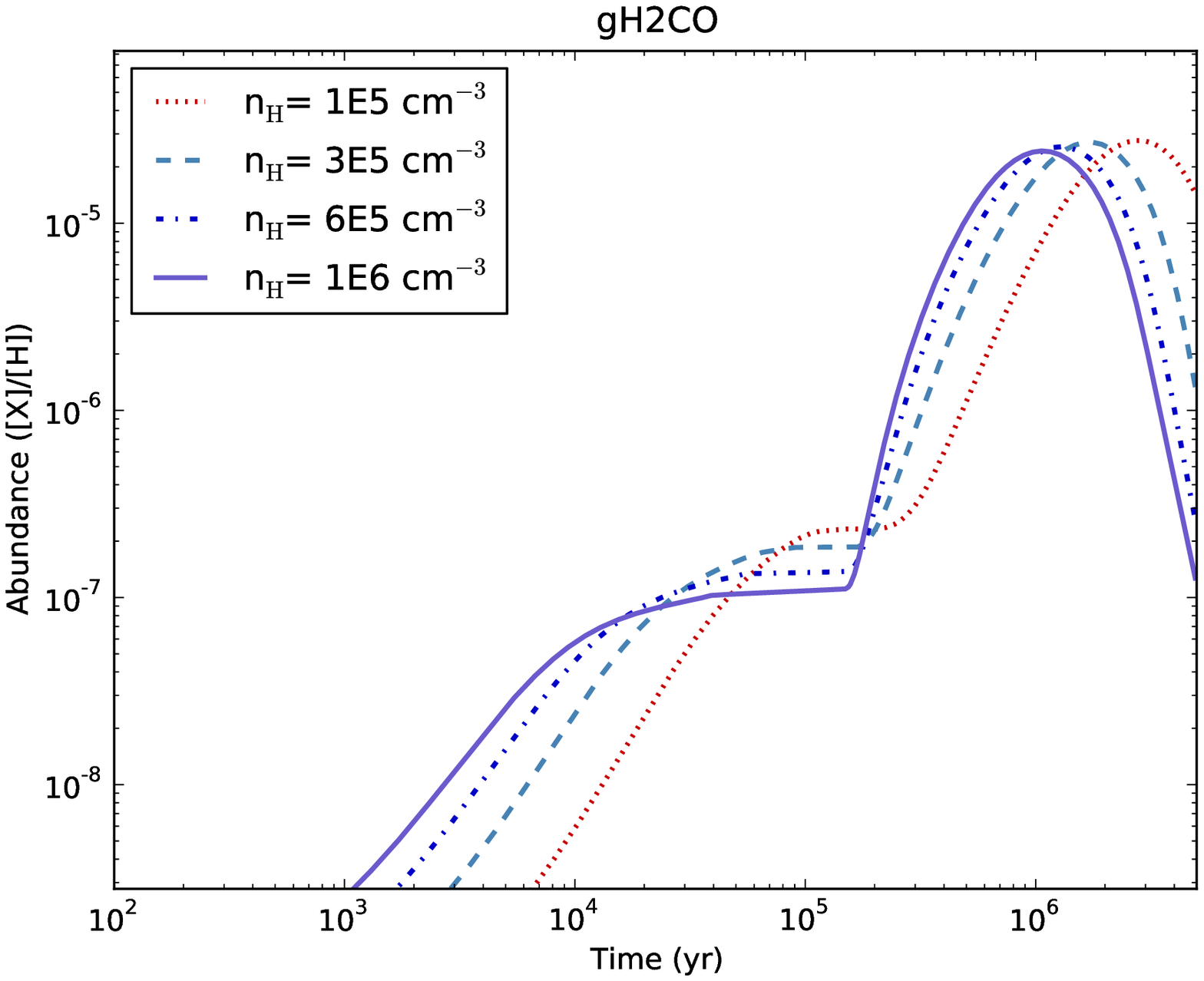}%
\includegraphics[width=0.49\textwidth]{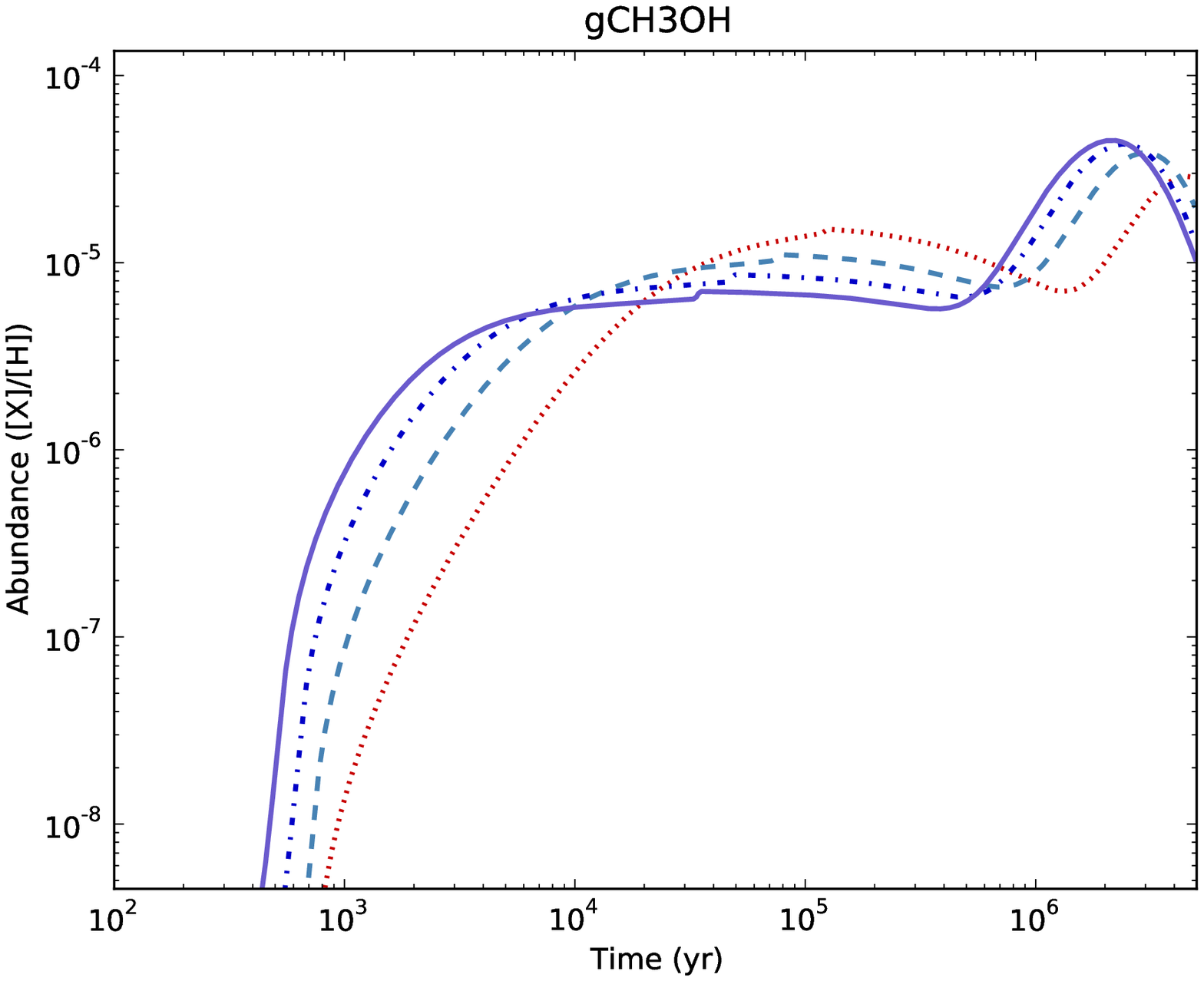}
\caption{The abundances of H$_2$CO and CH$_3$OH ice as a function of time for
different densities. The temperature is fixed to 21 K.}
\label{fig:togethernH}
\end{figure*}

Regarding the density dependence, Fig.~\ref{fig:vsnH} shows that the abundances
of the gas phase species at 6$\times$10$^5$ yr generally decrease with
increasing density, because the accretion of molecules onto the dust grains is
faster for a higher density.  This does not necessarily mean that the
abundances of the surface species always increase with density.  For example,
the abundances of CH$_4$ ice and CH$_3$OH ice decrease with higher density,
while the abundance of H$_2$CO ice has the opposite trend.  One important
factor is the time when we look at the system.  In Fig. \ref{fig:togethernH} we
plot the abundances of H$_2$CO ice and CH$_3$OH ice as a function of time
for different densities, while the temperature is fixed at 21 K. It can be seen
that at a given time, the abundances of H$_2$CO ice or CH$_3$OH ice can either
increase or decrease when the density is increased. We note that the evolution
curves of these species have a quasi-oscillatory feature.  For the same
species, the evolution curves have a similar shape for different densities,
except that with a lower density the evolution is slower and thus the curves
are shifted rightward.



\subsection{Discussions and limits of the model}
\label{sec:limmodel}

It can be seen from Fig.~\ref{fig:res1}, Fig.~\ref{fig:vsnH}, and
Fig.~\ref{fig:togetherT} that H$_2$O$_2$ can be produced with a rather high
abundance at certain evolution stages or with certain physical parameters,
which is frequently higher than the current observed value \citep{Bergman11b},
especially in the early to intermediate time ($\sim$10$^3$ -- 10$^5$ yr).  One
natural question to ask is why H$_2$O$_2$ has not been commonly detected in the
interstellar medium.  One possibility is that its spectral lines have been
overlooked in the past. Another simple explanation would be that it is
over-produced in our model, because in the past the chemistry of H$_2$O$_2$ (as
well as O$_2$H and O$_3$, etc.) may not have been studied in detail in the
astro-chemistry context (especially the destruction reactions in the gas
phase), so that its destruction routes might be incomplete. For example, there
is no reaction in which H$_2$O$_2$ is destroyed by reacting with H$_3^+$
in the current mainstream gas phase chemical networks. As a rough estimate,
if we take the rate of destruction by such a reaction to be the same as the
rate of the reaction ${\rm H_3^+ + H_2O\rightarrow H_3O^+ + H_2}$, then the
abundance of gas phase H$_2$O$_2$ can be reduced by about one order of
magnitude.  Further theoretical/experimental studies of the H$_2$O$_2$
chemistry would thus be very helpful, given the fact that it has been detected
recently and its potential important role in the grain chemistry of water. A
third possibility is that the rarity of H$_2$O$_2$ might be an age effect. From
Fig.~\ref{fig:togetherT} we notice that the abundance of H$_2$O$_2$ is very
high only in a relatively early stage (before $\sim$5$\times$10$^5$~yr). If for
certain reason most of the cloud cores being observationally studied are older
than this (due to some selection effects), then the H$_2$O$_2$ abundance in
these objects would be too low to detect. Basically, at least three physical
parameters, namely age, density, and temperature, are relevant. A probability
distribution of these three parameters of the cloud cores would help to give
the detection probability of H$_2$O$_2$ (and any other molecules).
In this sense, $\rho$ Oph A may be considered special in the sense that
it has a relatively high density ($\sim$10$^6$ cm$^{-3}$) and temperature (20
-- 30 K), while most dark clouds with a high density ($\gtrsim$10$^4$
cm$^{-3}$) tend to be very cold ($\lesssim$15 K) \citep{Bergin07}.
An inhomogeneous physical condition would make the situation more complex, which
may require a self-consistent dynamical-chemical model.  However, a thorough
study on these possibilities has to be left for future work.


On the other hand, for CH$_3$OH, although it has been studied quite extensively
in the past, we notice that the gas phase reactions associated with it contain
some important (although not decisive) differences between the OSU09 network
and the UMIST RATE06 network.  For example, the reaction between CH and
CH$_3$OH to form CH$_3$ and H$_2$CO has a rate
2.49$\times$10$^{-10}$$(T/300)^{-1.93}$ in the UMIST RATE06 network, but it
does not exist in the OSU09 network.  One possible problem is that the
temperature of cold interstellar medium (at most several tens of Kelvins) is
out of the indicated valid range for many reactions in RATE06, and it is not
clear how to extrapolate these reaction rates correctly, although we have
closely followed the instructions in \citet{Woodall07}.  In our modeling we
have been using the RATE06 network.

The energy barriers for the hydrogenation of CO and H$_2$CO on the
grain are both taken to be 2500 K. \citet{Woon02} calculated the barrier
heights of
these two reactions, giving a value of $\sim$2740 K and 3100 K, respectively,
in the case three water molecules are present, with zero point energy
corrections added. If these values were adopted, then the observed abundances
of H$_2$CO and CH$_3$OH can only be reproduced within one order of magnitude
at best.  However, \citet{Goumans11} gives a lower barrier height ($\sim$2200
K) for the hydrogenation of H$_2$CO, which would give a better agreement with
the observational results than if \citet{Woon02} were used in our model.

The chemical desorption is very important for the abundances of the gas phase
H$_2$O$_2$ and CH$_3$OH. However, the efficiency of this mechanism (the ``$a$''
parameter in \citet{Garrod07}) is uncertain. \citet{Garrod11} adopted a low
value of 0.01 for it, to avoid over-production of some gas phase species
\citep{Garrod07}. We use a value of $0.1$ in our study, and the abundances of
H$_2$CO and CH$_3$OH are not over-produced, except possibly in the early stages
of the evolution. We note that the temperature and density of major concern in
our study is around 20 K and 6$\times$10$^5$ cm$^{-3}$, while in
\citet{Garrod07} the temperature and density are set to 10 K and
2$\times$10$^4$ cm$^{-3}$, respectively. As a test we run our model with the
latter physical condition, and in this case CH$_3$OH and H$_2$CO are indeed
over-produced by about one order of magnitude.  Thus it seems that the
efficiency of chemical desorption depends on the temperature of the dust grain,
in the sense that at higher temperature the probability that the product of a
surface exothermic reaction gets ejected to the gas phase is higher.  However,
a detailed study of this possibility is out of the scope of the present paper.

Regarding the formation of water ice, \citet{Bergin98} proposed an interesting
mechanism in which water is first formed in the high temperature shocked gas,
and then gets adsorbed onto the dust grains in the post-shock phase.  This
mechanism may act as an alternative or supplement to the grain chemistry route.
It is not our aim here to discuss to what extent this mechanism contributes to
the water ice budget. However, we remark that even with a temperature 1000 --
2000 K, the amount of H$_2$O$_2$ produced in a pure gas phase chemistry (using the UMIST RATE06 network) is
still much lower than the detected level.

\section{Conclusions}
\label{sec:concl}

With a gas-grain chemical model which properly takes into account the
desorption of grain surface species by the heat released by chemical reactions,
we reproduce the observed abundances of H$_2$O$_2$ in $\rho$ Oph A at a time
of $\sim$6$\times$10$^5$ yr. The solid phase H$_2$O$_2$ abundance is
very low in this stage. However, a H$_2$O$_2$ to H$_2$O ratio of a few
percent might be obtained in the solid phase if the layered structure of grain
mantle is taken into account.

The abundances of other species such as H$_2$CO, CH$_3$OH, and O$_2$ detected
in the same object can also be reasonably reproduced at a time of
$\sim$6$\times$10$^5$ yr. Such a time scale is consistent with the evolution
time scale estimated through dynamical considerations.

O$_2$H is a precursor of H$_2$O$_2$ on the dust grain, and we predict that it
has a gas-phase abundance with the same order-of-magnitude of H$_2$O$_2$ and
should thus be detectable. Observational searches for it are under way.

For physical conditions relevant to $\rho$ Oph A, water is mainly in solid
form, being the dominant grain mantle material.  Its gas phase abundance is
only of the order of 10$^{-8}$ according to our model.

We note that the abundance of gas-phase H$_2$O$_2$ in our model results can be
much higher than the current observed level for a range of physical conditions.
This may suggest that its gas phase destructing channels are incomplete. Due to
the potential important role played by H$_2$O$_2$ in the formation of water,
its reaction network needs to be studied more thoroughly in the future.

Other uncertainties in our modeling include the ratio between the diffusion
energy barrier to the binding energy of a species on the grain surface, the
activation energy barriers of certain key reactions, as well as the efficiency
of the chemical desorption mechanism. In the present work we mainly make use
of their canonical values, or values which give good match to the
observational results, and we also vary them to see the effects on the
resulting abundances, which are significant in many cases.

\begin{acknowledgements}
We thank the anonymous referee and our editor Malcolm Walmsley for very detailed and constructive comments.
We also thank A. G. G. M. Tielens for interesting discussions.
F. Du and B. Parise are financially supported by the Deutsche
Forschungsgemeinschaft Emmy Noether program under grant PA1692/1-1.
\end{acknowledgements}

\bibliographystyle{aa}
\bibliography{../notes/references}

\begin{appendix}

\section{An explanation of the spike-like features in the evolution curves}
\label{app:expSpike}

\begin{figure}[htbp]
\centering
\includegraphics[width=0.49\textwidth]{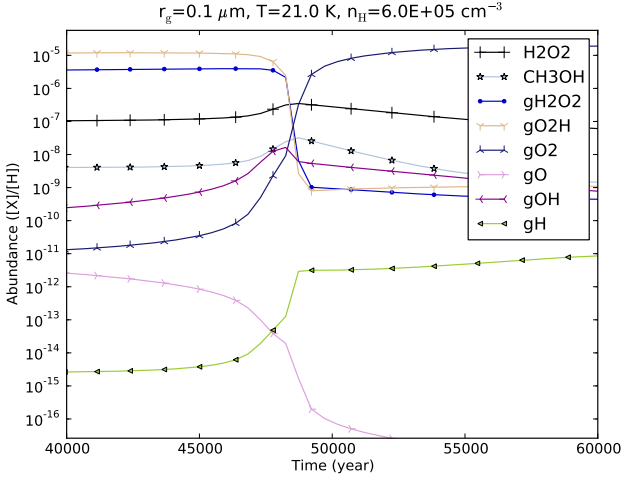}
\caption{A zoom-in plot of Fig. \ref{fig:res1} (with several species removed
and several added), focusing on the spike-like feature (though nothing looks
like a spike anymore). The abundance of gH has been multiplied by a factor of
10$^{12}$ (to make the plot look more compact).}
\label{fig:res1zoomin}
\end{figure}

In Fig. \ref{fig:res1} we note that at a time of $\sim$5$\times$10$^4$ yr a
spike-like feature appears in the evolution curves of some species (e.g. gas
phase H$_2$O$_2$ and CH$_3$OH), while the abundances of some other species
(e.g. H$_2$O$_2$ and O$_2$ on the grain) change very rapidly at the
same time.  While these features may appear to be caused by faults in the
program for solving the set of differential equations, however, after solving
the same problem with the Monte Carlo method which is immune to such
numerical instabilities, we find that these features are still present,
indicating that they are genuine. How could a smooth ordinary differential
equation system generate such an almost-singular feature? In Fig.
\ref{fig:res1zoomin} we make a zoom-in of Fig. \ref{fig:res1} (with several
species added and several species removed).  It can be seen that although the
time scale of the spike-like feature is relatively short, the evolution is
always smooth (except the discreteness caused by the finite sampling of the
curve).  Then what determines the appearance of such a feature?

As atomic hydrogen is central to the surface chemistry, we first look at the
most important reactions governing its abundance on the grain.
The main reactions consuming atomic hydrogen on the grains are (in the
following a species name preceded by a ``g'' means a species on the grain,
otherwise it is in the gas phase)
\begin{equation}
   {\rm gH + gO_2H} \xrightarrow{k_1} \left\{
     \begin{aligned}
       & {\rm gOH + gOH,} \\
       & {\rm gH_2O_2,}\\
       & {\rm OH + OH,} \\
       & {\rm H_2O_2,}
     \end{aligned}
   \right.
\end{equation}
\begin{equation}
  {\rm gH + gH_2O_2} \xrightarrow{k_2} \left\{
    \begin{aligned}
      & {\rm gH_2O + gOH},\\
      & {\rm H_2O + OH},
    \end{aligned}
  \right.
\end{equation}
while others such as the reactions with gO$_3$ and gCH$_2$OH are relatively
unimportant in the early times.
It is mainly produced by
\begin{equation}
  {\rm H \xrightarrow{k_3} gH,}
\end{equation}
\begin{equation}
  {\rm gOH + gCO \xrightarrow{k_4} gCO_2 + gH,}
\end{equation}
\begin{equation}
  {\rm gH_2O \xrightarrow{k_5} gOH + gH.}
\end{equation}

As the abundance of atomic H in the gas phase and the abundance of gCO and
gH$_2$O change relatively smoothly, they can be viewed as constant in a short
time scale.  The main reactions for the consumption and production of gOH are
\begin{equation}
  {\rm gO + gOH} \xrightarrow{k_6} \left\{
    \begin{aligned}
      &{\rm gO_2H,} \\
      &{\rm O_2H,} \\
    \end{aligned}
  \right.
  \label{eq:gOgOH}
\end{equation}
\begin{equation}
  {\rm gCO + gOH} \xrightarrow{k_7} \left\{
    \begin{aligned}
      &{\rm gCO_2 + gH,} \\
      &{\rm CO_2 + H,} \\
    \end{aligned}
  \right.
  \label{eq:gCOgOH}
\end{equation}
\begin{equation}
   {\rm gH + gO_2H} \xrightarrow{k_8}
       {\rm gOH + gOH,}
  \label{eq:gHgO2H}
\end{equation}
\begin{equation}
  {\rm gH + gH_2O_2} \xrightarrow{k_9}
      {\rm gH_2O + gOH},
\end{equation}
\begin{equation}
  {\rm gH_2O} \xrightarrow{k_{10}} {\rm gH+gOH}.
\end{equation}

From the above reaction list we may write the evolution equation of gH and gOH
as\footnote{Here for brevity we use the name of a species to denote its average population on a
single grain; for example, if gCO=100, it mains on average there are 100 CO
molecules on a single grain.}
\begin{align}
  \partial_t{\rm gH} = & -k_1 {\rm gH gO_2H} - k_2 {\rm gH gH_2O_2} \nonumber \\
                       & +k_3 {\rm H} + k_4 {\rm gOH gCO} + k_5 {\rm gH_2O},
  \label{eq:gHEq}
\end{align}
\begin{align}
  \partial_t{\rm gOH} = & -k_6 {\rm gO gOH} - k_7 {\rm gCO gOH} \nonumber \\
                        & +2k_8 {\rm gH gO_2H} + k_9 {\rm gH gH_2O_2} + k_{10}{\rm gH_2O},
  \label{eq:gOHEq}
\end{align}
or in a more succinct form
\begin{align}
  \partial_t{\rm gH} = & \kappa_1 {\rm gH} + \kappa_2 {\rm gOH} + b_1, \nonumber \\
  \partial_t{\rm gOH} = & \kappa_3 {\rm gH} + \kappa_4 {\rm gOH} + b_2,
  \label{eq:gHgOH}
\end{align}
where
\begin{align*}
  \kappa_1 =& -k_1{\rm gO_2H} - k_2{\rm gH_2O_2},\\
  \kappa_2 =& k_4{\rm gCO},\\
  \kappa_3 =& 2k_8{\rm gO_2H}+k_9{\rm gH_2O_2},\\
  \kappa_4 =& -k_6{\rm gO}-k_7{\rm gCO},\\
  b_1 =& k_3{\rm H} + k_5{\rm gH_2O},\\
  b_2 =& k_{10}{\rm gH_2O}.
\end{align*}

If we view $\kappa_1$ -- $\kappa_4$ as well as $b_1$ and $b_2$ as constants (of
course they are not), then equation (\ref{eq:gHgOH}) can be solved exactly; the
solution contains an exponential part and a constant part. The amplitude of the
exponential part will be inversely proportional to the determinant of the
coefficient matrix
$${\rm det}(\kappa) = 
\left(
\begin{array}{cc}
\kappa_1 & \kappa_2\\
\kappa_3 & \kappa_4
\end{array}
\right).
$$
As $\kappa_1$ -- $\kappa_4$ are not really constant in our problem, we expect
that when they become such that det($\kappa$) is close to zero, a spike-like or
jump-like behavior would appear. Namely, we require
\begin{align}
  \frac{\kappa_1\kappa_4}{\kappa_2\kappa_3} 
  =&\frac{(k_1{\rm gO_2H} + k_2{\rm gH_2O_2})(k_6{\rm gO}+k_7{\rm gCO})}
        {(k_4{\rm gCO})(2k_8{\rm gO_2H}+k_9{\rm gH_2O_2})} \nonumber \\
  \simeq&\frac{{\rm gO_2H/gH_2O_2 + 1}}{1.2{\rm gO_2H/gH_2O_2}+1}
        \times \left(2.5\times10^8{\rm gO/gCO + 1}\right) \nonumber \\
  \simeq& 1.
  \label{eq:spikecondition}
\end{align}  
In the second line of the above equation the actual value of the parameters
have been inserted. These parameters depend on the physical conditions.



\begin{figure}[htbp]
\centering
\includegraphics[width=0.49\textwidth]{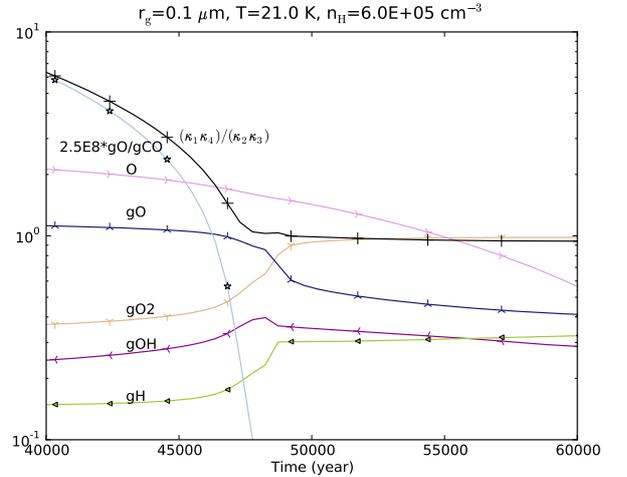}
\caption{The ratio $(\kappa_1\kappa_4)/(\kappa_2\kappa_3)$ and the value of
2.5$\times$10$^8$$\times$gO/gCO as a function of time. They are to scale.  The
abundances of several other species are plotted only for reference (not to
scale).}
\label{fig:gOgCO}
\end{figure}

To satisfy this condition (at least approximately), gO/gCO should be very
small.  In Fig. \ref{fig:gOgCO} the ratio
$(\kappa_1\kappa_4)/(\kappa_2\kappa_3)$ and the value of
2.5$\times$10$^8$gO/gCO are plotted as a function of time. The abundances of O,
gO, gO$_2$, gH, and gOH are also plotted for reference (not to scale). It can
be seen that the $(\kappa_1\kappa_4)/(\kappa_2\kappa_3)$ ratio does decrease
and approach a value of unity before the time of the spike/jump feature, and
the gO/gCO ratio does drop to a very low value monotonically.

The above mathematical argument can also be understood intuitively. When the gO
abundance is so low that reaction (\ref{eq:gOgOH}) can be neglected, gOH is
only destroyed by reaction (\ref{eq:gCOgOH}). Each time a gOH radical is
consumed, one gH is created (if we neglect the desorption process), and this gH
will quickly react with gH$_2$O$_2$ or gO$_2$H to create one or two gOHs. So
there will be a net gain in the gOH abundance, leading to its fast growth, and
the gH abundance will increase accordingly.  Thus we see that reaction
(\ref{eq:gHgO2H}) is crucial in that it produced two gOH radicals by consuming
only one gH.

Therefore for such a spike-like feature to occur, the abundance of gO must
decrease to a low value such that the reaction between gOH and gO becomes
unimportant in consuming gOH.  Namely, we require ${\rm [gO] \lesssim [gCO]\
k_7/k_6}\simeq 5\times10^{-13}$.  The abundances of atomic oxygen in the gas
phase and on the grain surface are related by ${\rm O/gO \simeq k_{\rm
evap}(gO)/k_{\rm ad}(O)\simeq3\times10^{6}}$, so equivalently we require ${\rm
[O]<1.5\times10^{-6}}$ (i.e. about a factor of 200 less than the initial O
abundance) at the time of the spike-like feature.  Atomic oxygen is mainly
consumed on the grain surface by reacting with another gO to form gO$_2$ (for
$t\lesssim10^3$ yr) or by reacting with gOH to form gO$_2$H.  Only the latter
is relevant here.  As the abundance of gOH does not change much before the
spike-like feature, the time scale for the consumption of atomic oxygen can be
estimated to be ${\ln200\times({\rm O/gO})/(k_6{\rm gOH})}\simeq10^5$~yr, which
is of the same order of magnitude of the time of occurrence of the spike-like
feature.



The time scale for the endurance of the spike-like feature itself can be
estimated to be the time scale for the exhaustion of gO$_2$H or gH$_2$O$_2$ (so
that equations (\ref{eq:gHEq}) and (\ref{eq:gOHEq}) do not hold anymore) by
reacting with gH, which is $$1/(k_1{\rm gH})\simeq 10^3\ {\rm yr},$$ where the
gH population (the average number of atomic H on a single grain) is taken to be
a median value (10$^{-14}$) during the rapidly-varying period.
The fact that equation (\ref{eq:spikecondition}) seems to hold after this
period (see Fig. \ref{fig:gOgCO}) does not mean that gH will keep increasing
rapidly, simply because the premise of our argument, namely equations
(\ref{eq:gHEq}) and (\ref{eq:gOHEq}) are not a good description of the
evolution of gH and gOH anymore.

As the abundance of H on the grain increases, almost all the O$_3$ on the grain
are converted into O$_2$ and OH. O$_2$ molecules on the grain are then consumed
by the slower reaction ${\rm gH+gO_2\rightarrow gO_2H}$, with a time scale
$\sim$10$^5$ yr. This explains the prominent peak in the evolution curve of
gO$_2$ (see Fig. \ref{fig:res1}).

Would the spike-like feature have any practical significance (especially
observationally)? Ideally, such a short-time feature may be used to contrain
the age of a dense cloud, by discrimiating the abundances of certain species
between their early-time values (before the spike-like feature) and late-time
values (after the spike-like feature). However, due to its dependence on the
reaction network being used, which usually contains a lot of uncertainties and
is subject to change when new experiments are carried out, the question whether
this feature is really relevant for the study of interstellar medium can only
be answered by future investigations.

\Online

\pagebreak
\section{The surface reaction network used in this work}
\label{app:surfacenetwork}
\tablebib{
HHL92: \citet{Hasegawa92}; 
ICet10: \citet{Ioppolo10};
CIRL10: \citet{Cuppen10};
ICet08: \citet{Ioppolo08};
G11: \citet{Goumans11};
FCet09: \citet{Fuchs09}
GWH08: \citet{Garrod08a};
AR77: \citet{Allen77};
TH82: \citet{Tielens82};
GMet08: \citet{Goumans08};
ABet04: \citet{Atkinson04};
RH00: \citet{Ruffle00};
RH01: \citet{Ruffle01a};
}
\longtab{1}{
\begin{longtable}{r l c c l}
\caption{The surface network used in this work. The photo-dissociation
reactions induced by cosmic rays and the chemical desorption reactions are not
included here.}\\
\hline\hline
Num  & Reaction             & Branching ratio & Energy barrier (K) & Reference \\
\hline
\endfirsthead
\caption{continued.}\\
\hline\hline
Num  & Reaction             & Branching ratio & Energy barrier (K) & Reference \\
\hline
\endhead
\hline
\endfoot
  1  &  H     +    H           $\rightarrow$     H$_2$                  & 1.0    &    0.0  &  HHL92  \\
  2  &  H     +    O           $\rightarrow$     OH                  & 1.0    &    0.0  &  ICet10  \\
  3  &  H     +    O$_2$          $\rightarrow$     O$_2$H                 & 1.0    &  600.0  &  Estimated  \\
  4  &  H     +    O$_3$          $\rightarrow$     O$_2$      +  OH       & 1.0    &  200.0  &  Estimated  \\
  5  &  H     +    OH          $\rightarrow$     H$_2$O                 & 1.0    &    0.0  &  ICet10  \\
  6  &  H     +    O$_2$H         $\rightarrow$     H$_2$O$_2$                & 0.38   &    0.0  &  CIRL10  \\
  7  &  H     +    O$_2$H         $\rightarrow$     OH      +  OH       & 0.62   &    0.0  &  CIRL10  \\
  8  &  H     +    H$_2$O$_2$        $\rightarrow$     H$_2$O     +  OH       & 1.0    &    0.0  &  ICet08  \\
  9  &  H     +    CO          $\rightarrow$     HCO                 & 0.5    & 2500.0  &  GWH08  \\
 10  &  H     +    HCO         $\rightarrow$     H$_2$CO                & 1.0    &    0.0  &  FCet09  \\
 11  &  H     +    H$_2$CO        $\rightarrow$     CH$_3$O                & 0.5    & 2500.0  &  RH00  \\
 12  &  H     +    H$_2$CO        $\rightarrow$     HCO     +  H$_2$       & 0.5    & 3000.0  &  G11  \\
 13  &  H     +    CH$_3$O        $\rightarrow$     CH$_3$OH               & 1.0    &    0.0  &  FCet09  \\
 14  &  H     +    CH$_2$OH       $\rightarrow$     CH$_3$OH               & 1.0    &    0.0  &  GWH08  \\
 15  &  H     +    HCOO        $\rightarrow$     HCOOH               & 1.0    &    0.0  &  AR77  \\
 16  &  H     +    C           $\rightarrow$     CH                  & 1.0    &    0.0  &  AR77  \\
 17  &  H     +    CH          $\rightarrow$     CH$_2$                 & 1.0    &    0.0  &  AR77  \\
 18  &  H     +    CH$_2$         $\rightarrow$     CH$_3$                 & 1.0    &    0.0  &  AR77  \\
 19  &  H     +    CH$_3$         $\rightarrow$     CH$_4$                 & 1.0    &    0.0  &  AR77  \\
 20  &  H     +    N           $\rightarrow$     NH                  & 1.0    &    0.0  &  AR77  \\
 21  &  H     +    NH          $\rightarrow$     NH$_2$                 & 1.0    &    0.0  &  AR77  \\
 22  &  H     +    NH$_2$         $\rightarrow$     NH$_3$                 & 1.0    &    0.0  &  AR77  \\
 23  &  H     +    S           $\rightarrow$     HS                  & 1.0    &    0.0  &  HHL92  \\
 24  &  H     +    HS          $\rightarrow$     H$_2$S                 & 1.0    &    0.0  &  HHL92  \\
 25  &  H     +    H$_2$S         $\rightarrow$     HS      +  H$_2$       & 1.0    &  860.0  &  TH82  \\
 26  &  H     +    CS          $\rightarrow$     HCS                 & 1.0    &    0.0  &  HHL92  \\
 27  &  C     +    S           $\rightarrow$     CS                  & 1.0    &    0.0  &  HHL92  \\
 28  &  O     +    S           $\rightarrow$     SO                  & 1.0    &    0.0  &  HHL92  \\
 29  &  O     +    SO          $\rightarrow$     SO$_2$                 & 1.0    &    0.0  &  HHL92  \\
 30  &  O     +    CS          $\rightarrow$     OCS                 & 1.0    &    0.0  &  HHL92  \\
 31  &  H     +    CN          $\rightarrow$     HCN                 & 1.0    &    0.0  &  AR77  \\
 32  &  H     +    NO          $\rightarrow$     HNO                 & 1.0    &    0.0  &  AR77  \\
 33  &  H     +    NO$_2$         $\rightarrow$     HNO$_2$                & 1.0    &    0.0  &  AR77  \\
 34  &  H     +    NO$_3$         $\rightarrow$     HNO$_3$                & 1.0    &    0.0  &  AR77  \\
 35  &  H     +    N$_2$H         $\rightarrow$     N$_2$H$_2$                & 1.0    &    0.0  &  AR77  \\
 36  &  H     +    N$_2$H$_2$        $\rightarrow$     N$_2$H     +  H$_2$       & 1.0    &  650.0  &  HHL92  \\
 37  &  H     +    NHCO        $\rightarrow$     NH$_2$CO               & 1.0    &    0.0  &  AR77  \\
 38  &  H     +    NH$_2$CO       $\rightarrow$     NH$_2$CHO              & 1.0    &    0.0  &  AR77  \\
 39  &  N     +    HCO         $\rightarrow$     NHCO                & 1.0    &    0.0  &  AR77  \\
 40  &  CH    +    CH          $\rightarrow$     C$_2$H$_2$                & 1.0    &    0.0  &  HHL92  \\
 41  &  O     +    O           $\rightarrow$     O$_2$                  & 1.0    &    0.0  &  AR77  \\
 42  &  O     +    O$_2$          $\rightarrow$     O$_3$                  & 1.0    &    0.0  &  ABet04  \\
 43  &  O     +    CO          $\rightarrow$     CO$_2$                 & 1.0    & 1580.0  &  GMet08  \\
 44  &  O     +    HCO         $\rightarrow$     HCOO                & 0.5    &    0.0  &  GMet08  \\
 45  &  O     +    HCO         $\rightarrow$     CO$_2$     +  H        & 0.5    &    0.0  &  GMet08  \\
 46  &  O     +    N           $\rightarrow$     NO                  & 1.0    &    0.0  &  AR77  \\
 47  &  O     +    NO          $\rightarrow$     NO$_2$                 & 1.0    &    0.0  &  AR77  \\
 48  &  O     +    NO$_2$         $\rightarrow$     NO$_3$                 & 1.0    &    0.0  &  AR77  \\
 49  &  O     +    CN          $\rightarrow$     OCN                 & 1.0    &    0.0  &  AR77  \\
 50  &  C     +    N           $\rightarrow$     CN                  & 1.0    &    0.0  &  AR77  \\
 51  &  N     +    N           $\rightarrow$     N$_2$                  & 1.0    &    0.0  &  AR77  \\
 52  &  N     +    NH          $\rightarrow$     N$_2$H                 & 1.0    &    0.0  &  AR77  \\
 53  &  H$_2$    +    OH          $\rightarrow$     H$_2$O     +  H        & 1.0    & 2100.0  &  ABet04  \\
 54  &  OH    +    CO          $\rightarrow$     CO$_2$     +  H        & 1.0    &   80.0  &  RH01  \\
 55  &  H     +    C$_2$          $\rightarrow$     C$_2$H                 & 1.0    &    0.0  &  HHL92  \\
 56  &  H     +    N$_2$          $\rightarrow$     N$_2$H                 & 1.0    & 1200.0  &  HHL92  \\
 57  &  H     +    C$_2$H         $\rightarrow$     C$_2$H$_2$                & 1.0    &    0.0  &  HHL92  \\
 58  &  H     +    HOC         $\rightarrow$     CHOH                & 1.0    &    0.0  &  HHL92  \\
 59  &  C     +    OH          $\rightarrow$     HOC                 & 0.5    &    0.0  &  HHL92  \\
 60  &  C     +    OH          $\rightarrow$     CO      +  H        & 0.5    &    0.0  &  HHL92  \\
 61  &  CH    +    OH          $\rightarrow$     CHOH                & 1.0    &    0.0  &  HHL92  \\
 62  &  H     +    CHOH        $\rightarrow$     CH$_2$OH               & 1.0    &    0.0  &  HHL92  \\
 63  &  OH    +    OH          $\rightarrow$     H$_2$O$_2$                & 1.0    &    0.0  &  HHL92  \\
 64  &  OH    +    CH$_2$         $\rightarrow$     CH$_2$OH               & 1.0    &    0.0  &  HHL92  \\
 65  &  C     +    C           $\rightarrow$     C$_2$                  & 1.0    &    0.0  &  HHL92  \\
 66  &  C     +    O$_2$          $\rightarrow$     CO      +  O        & 1.0    &    0.0  &  HHL92  \\
 67  &  O     +    CH          $\rightarrow$     HCO                 & 1.0    &    0.0  &  HHL92  \\
 68  &  O     +    OH          $\rightarrow$     O$_2$H                 & 1.0    &    0.0  &  HHL92  \\
 69  &  O     +    CH$_2$         $\rightarrow$     H$_2$CO                & 1.0    &    0.0  &  HHL92  \\
 70  &  O     +    CH$_3$         $\rightarrow$     CH$_2$OH               & 1.0    &    0.0  &  HHL92  \\
 71  &  C     +    O           $\rightarrow$     CO                  & 1.0    &    0.0  &  HHL92  \\
 72  &  C     +    CH          $\rightarrow$     C$_2$H                 & 1.0    &    0.0  &  HHL92  \\
 73  &  C     +    NH          $\rightarrow$     HNC                 & 1.0    &    0.0  &  HHL92  \\
 74  &  C     +    CH$_2$         $\rightarrow$     C$_2$H$_2$                & 1.0    &    0.0  &  HHL92  \\
 75  &  C     +    NH$_2$         $\rightarrow$     HNC     +  H        & 1.0    &    0.0  &  HHL92  \\
 76  &  N     +    CH          $\rightarrow$     HCN                 & 1.0    &    0.0  &  HHL92  \\
 77  &  N     +    NH$_2$         $\rightarrow$     N$_2$H$_2$                & 1.0    &    0.0  &  HHL92  \\
 78  &  O     +    NH          $\rightarrow$     HNO                 & 1.0    &    0.0  &  HHL92  \\
\hline
\end{longtable}
}

\pagebreak
\section{The enthalpies of the surface species}
\label{app:speciesenthalpy}
\tablebib{
BM02: \citet{Binnewies99};
NIST Webbook: http://webbook.nist.gov/chemistry/;
NCet10: \citet{Nagy10};
K98: \citet{Kaiser98};
VS91: \citet{Vandooren91};
Url1: http://chem.engr.utc.edu/webres/331f/teams-98/chp/Boiler\%20DC/tsld007.htm 
}
\longtab{1}{
\begin{longtable}{r l c l}
\caption{Enthalpies of the surface species considered in this work. They are
used to calculate the exoergicities of the surface reactions.}\\
\hline\hline
 Num & Species  & Enthalpy (kJ/mol)  & Reference \\
\hline
\endfirsthead
\caption{continued.}\\
\hline\hline
 Num & Species  & Enthalpy (kJ/mol)  & Reference \\
\hline
\endhead
\hline
\endfoot
\hline
  1  &  C          & 716.7  & NIST Webbook  \\
  2  &  CH         & 594.1  & BM02, p238  \\
  3  &  CH$_2$        & 386.4  & BM02, p240  \\
  4  &  CH$_3$        & 145.7  & BM02, p241  \\
  5  &  CH$_3$O       & 17.0   & NIST Webbook  \\
  6  &  CH$_2$OH      & -9.0   & NIST Webbook  \\
  7  &  CH$_3$OH      & -201.2 & BM02, p241  \\
  8  &  CH$_4$        & -74.9  & BM02, p241  \\
  9  &  CN         & 435.1  & BM02, p247  \\
 10  &  CO         & -110.5 & BM02, p251  \\
 11  &  CO$_2$        & -393.5 & BM02, p251  \\
 12  &  CS         & 280.3  & BM02, p253  \\
 13  &  H          & 218.0  & BM02, p558  \\
 14  &  H$_2$         & 0.0    & BM02, p568  \\
 15  &  H$_2$CO       & -115.9 & BM02, p240  \\
 16  &  H$_2$O        & -241.8 & BM02, p571  \\
 17  &  H$_2$O$_2$       & -135.8 & BM02, p572  \\
 18  &  H$_2$S        & -20.5  & BM02, p574  \\
 19  &  HCN        & 135.1  & BM02, p239  \\
 20  &  HNC        & 135.1  & Estimated  \\
 21  &  HCO        & 43.5   & BM02, p239  \\
 22  &  HCOO       & -386.8 & Url1  \\
 23  &  HCOOH      & -378.6 & BM02, p240  \\
 24  &  HCS        & 296.2  & K98  \\
 25  &  HNO        & 99.6   & BM02, p563  \\
 26  &  HNO$_2$       & -78.8  & BM02, p563  \\
 27  &  HNO$_3$       & -134.3 & BM02, p563  \\
 28  &  HS         & 139.3  & BM02, p567  \\
 29  &  N          & 472.7  & BM02, p693  \\
 30  &  N$_2$         & 0.0    & BM02, p699  \\
 31  &  N$_2$H        & 245.2  & VS91  \\
 32  &  N$_2$H$_2$       & 213.0  & BM02, p570  \\
 33  &  NH         & 376.6  & BM02, p563  \\
 34  &  NH$_2$        & 190.4  & BM02, p570  \\
 35  &  NH$_2$CHO     & -186.0 & NIST Webbook  \\
 36  &  NH$_2$CO      & -13.1  & NCet10  \\
 37  &  NH$_3$        & -45.9  & BM02, p577  \\
 38  &  NHCO       & -101.7 & BM02, p239  \\
 39  &  NO         & 90.3   & BM02, p695  \\
 40  &  NO$_2$        & 33.1   & BM02, p695  \\
 41  &  NO$_3$        & 71.1   & BM02, p695  \\
 42  &  O          & 249.2  & BM02, p733  \\
 43  &  O$_2$         & 0.0    & BM02, p741  \\
 44  &  O$_2$H        & 2.1    & BM02, p567  \\
 45  &  O$_3$         & 142.7  & BM02, p752  \\
 46  &  OCN        & 159.4  & BM02, p248  \\
 47  &  OCS        & -138.4 & BM02, p251  \\
 48  &  OH         & 39.0   & BM02, p566   \\
 49  &  S          & 277.0  & BM02, p811  \\
 50  &  SO         & 5.0    & BM02, p736  \\
 51  &  SO$_2$        & -296.8 & BM02, p743  \\
 52  &  C$_2$         & 837.74 & NIST Webbook  \\
 53  &  C$_2$H        & 476.98 & NIST Webbook  \\
 54  &  C$_2$H$_2$       & 226.73 & NIST Webbook  \\
\hline
\end{longtable}
}

\end{appendix}

\end{document}